\pdfoutput=1
\documentclass[%
prx,twocolumn,
notitlepage,
superscriptaddress,
 amsmath,amssymb,
 aps,
prx,
floatfix,
longbibliography
]{revtex4-1}
\usepackage[T1]{fontenc}

\usepackage{color}
\usepackage{amsmath}
\usepackage{amssymb}
\usepackage{bm}
\usepackage{graphicx}
\usepackage[usenames,dvipsnames]{xcolor}
\usepackage[%
  colorlinks=true,
  urlcolor=blue,
  linkcolor=blue,
  citecolor=blue
]{hyperref}

\usepackage{array}

\definecolor{bleu}{rgb}{0.16,0.2.5,0.36}

\newcommand{\EDF}{Appendix Fig.}

\begin{document}
\pdfoutput=1
\title{Non-orientable  order and non-Abelian response in  frustrated metamaterials}

\author{Xiaofei Guo}
\affiliation{Harbin Institute of Technology, China}
\affiliation{Institute of Physics, Universiteit van Amsterdam, 1098 XH Amsterdam, The Netherlands}
\author{Marcelo Guzmán}
\affiliation{Univ. Lyon, ENS de Lyon, Univ. Claude Bernard, CNRS, Laboratoire de Physique, F-69342, Lyon, France}
\author{David Carpentier}
\affiliation{Univ. Lyon, ENS de Lyon, Univ. Claude Bernard, CNRS, Laboratoire de Physique, F-69342, Lyon, France}
\author{Denis Bartolo}
\affiliation{Univ. Lyon, ENS de Lyon, Univ. Claude Bernard, CNRS, Laboratoire de Physique, F-69342, Lyon, France}
\author{Corentin Coulais}
\affiliation{Institute of Physics, Universiteit van Amsterdam, 1098 XH Amsterdam, The Netherlands}

\date{\today}
\maketitle

{\bf 
From atomic crystals to bird flocks, most forms of order are captured by the concept of spontaneous symmetry breaking~\cite{Chaikin1995,CrossHohenberg,Toner2005}.
This paradigm was challenged  by the discovery of topological order, 
in materials where the number of accessible states is not solely determined by the number of broken symmetries, but also by space topology~\cite{Wen_2017}. 
Until now however, the concept of topological order has been linked to quantum entanglement and has therefore remained out of reach in classical systems.
 Here, we show that classical systems whose global geometry frustrates the emergence of homogeneous order realise an unanticipated form of topological order defined by  non-orientable order-parameter bundles: non-orientable order.
We validate experimentally and theoretically this concept by designing frustrated mechanical metamaterials that spontaneously break a discrete symmetry under homogeneous load. 
While conventional order leads to a discrete ground-state degeneracy, we show that non-orientable order implies an extensive ground-state degeneracy---in the form of topologically protected zero-nodes and zero-lines. Our metamaterials escape the traditional classification of order by symmetry breaking. 
Considering more general stress distributions, we leverage non-orientable order to  engineer robust mechanical memory~\cite{Treml_PNAS2018,Nagel_review2019,Andres_dome2020,Chen_Nature2021,Yair_ice2021,Jules_arxiv2021,Bense2021} and achieve non-Abelian mechanical responses that carry an imprint of the braiding of local loads~\cite{Fruchart_Nature2020,Mietke2020}.
We envision this principle to open the way to designer materials that can robustly process information across multiple areas of physics, from mechanics to photonics and magnetism.}

\begin{figure*}[t!]
    \centering
    \includegraphics[width=\textwidth]{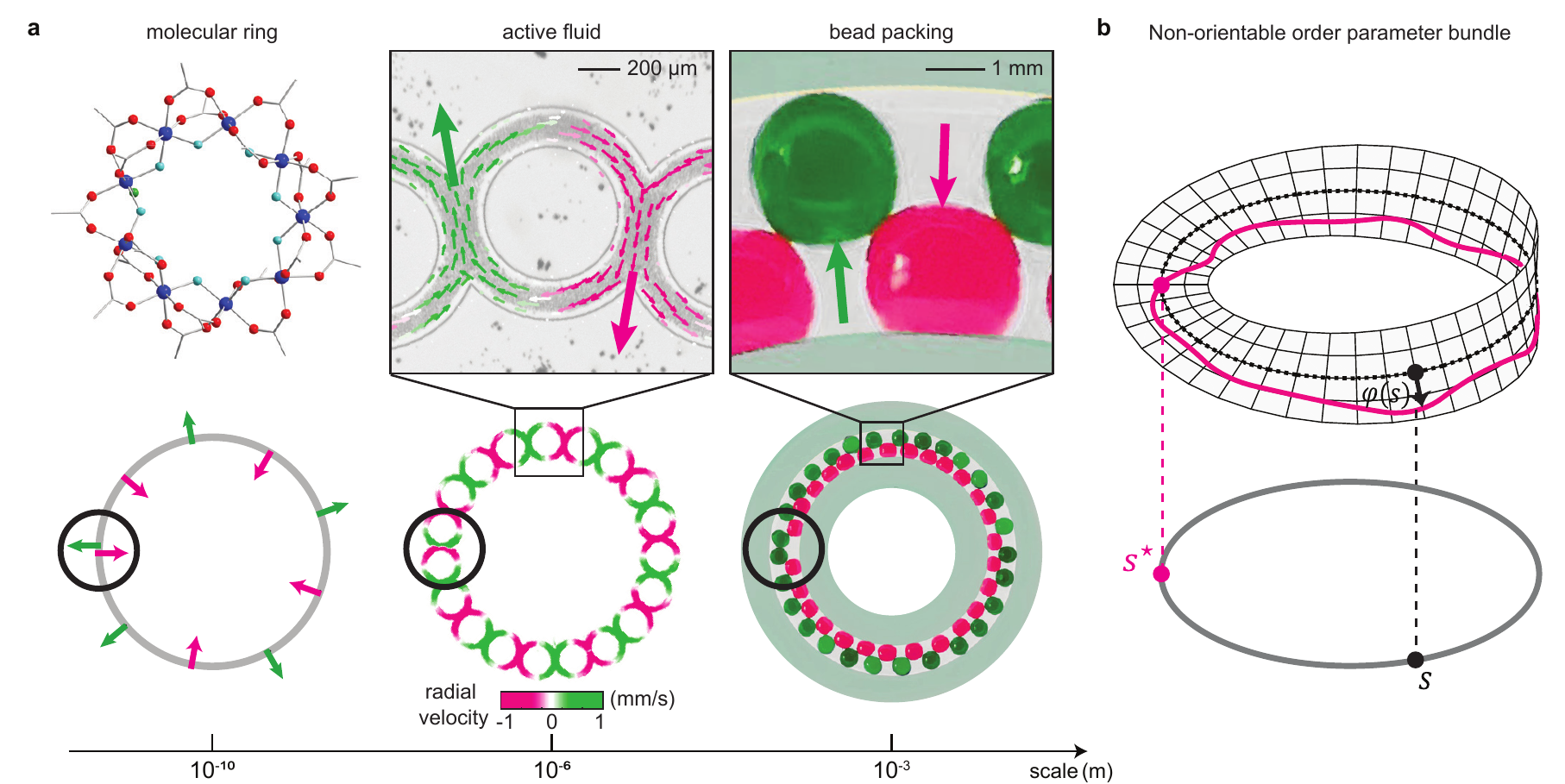}
    \caption{{\bf Frustration-induced non-orientable order.} %\xg{version1} 
    {\bf (a)} Three examples of globally frustrated systems spanning seven orders of magnitude in scale. 
    (Left): 
    A molecular antiferromagnetic ring made of a $\{$Cr$_9\}$ chemical compound (CCDC 893835), whose magnetic moments are located on $9$ Cr ions (blue) cannot accommodate a global antiferromagnetic order, see~\cite{Baker_PNAS2012}. 
    (Middle): 
    An active-fluid metamaterial made of $\sim 5~\mu \rm m$ large  colloidal rollers flowing at constant speed in 19 connected annular microchannels. The flow direction at the junctions of the ``fluidic gears'' self-organises to feature antiferromagnetic order. This order is globally frustrated when the number of annuli is odd, as a result the active flow vanishes in one unspecified junction (dark circle). This global frustration is clearly seen in the map of  radial component of the flow velocity $\mathbf v(\mathbf r)\cdot \mathbf r/r$, where the origin is set at the centre of the of the metastructure.  See also Supplementary Video 1. The experimental methods are thoroughly described in~\cite{Chardac2021}. 
    (Right):
    2 mm large polymer beads packed in a ring-shape channel. When confined in this quasi-one dimensional geometry, they realise a prototypical example of antiferromagnetic order (spheres are associated to $+1/2$ or $-1/2$ spin values if in contact with the outer or inner walls, respectively), see~\cite{Han2008} for a detailed analogy. When the ring perimeter is incommensurate with the length of the zigzag pattern antiferromagnetic order is globally frustrated (dark circle).
    {\bf (b)} The primary goal of our article is to quantitatively establish that the order parameters of globally frustrated physical systems define  non-orientable real fibre bundles. In the basic case  of frustrated 1D antiferromagnetic rings the order parameter field belongs to a M\"obius-strip bundle. In this non-trivial bundle, all sections $\varphi(s)$ (i.e. all field configurations) vanish at least once along the base circle at an unspecified point $s^\star$. The M\"obius strip is topologically distinct from the cylinder bundle associated to unfrustrated 1D antiferromagnets.
    }
    \label{fig:0}
\end{figure*}

Frustration arises whenever geometry is incompatible with the symmetries of local interactions~\cite{Toulouse1987,Moessner2006}. 
It materialises in equilibrium and out-of-equilibrium systems as diverse as electronic and synthetic spin ice~\cite{Balents2010,Nisoli2013}, active flow networks~\cite{Wioland2016,Wu2017,Forrow2017}, colloidal matter~\cite{Han2008,Ortiz2016,Molina2021}, and mechanical metamaterials~\cite{Bertoldi2017,Bertoldi_PRL2014,Celli_softmatter2018,Martin_NatPhys2020,Deng_PNAS2020,Yair_ice2021}.
In Fig.~\ref{fig:0}a,  we show three experiments that span seven orders of magnitude in size where local constraints promote an antiferromagnetic order. 
They all highlight the concept of global frustration: the existence of a loop that lassos the whole system, and along which  local constraints cannot be all satisfied.
In this paper, we combine mechanical experiments and theory to reveal that globally frustrated matter realises an uncharted class of classical topological order: non-orientable order. 
It is defined by a local symmetry breaking that is associated to a non-orientable order parameter bundle (Fig.~\ref{fig:0}b).    
We demonstrate that non-orientable order is characterised by an enriched ground-state degeneracy and features non-Abelian response.

\begin{figure*}[t!]
    \centering
    \includegraphics[width=\textwidth]{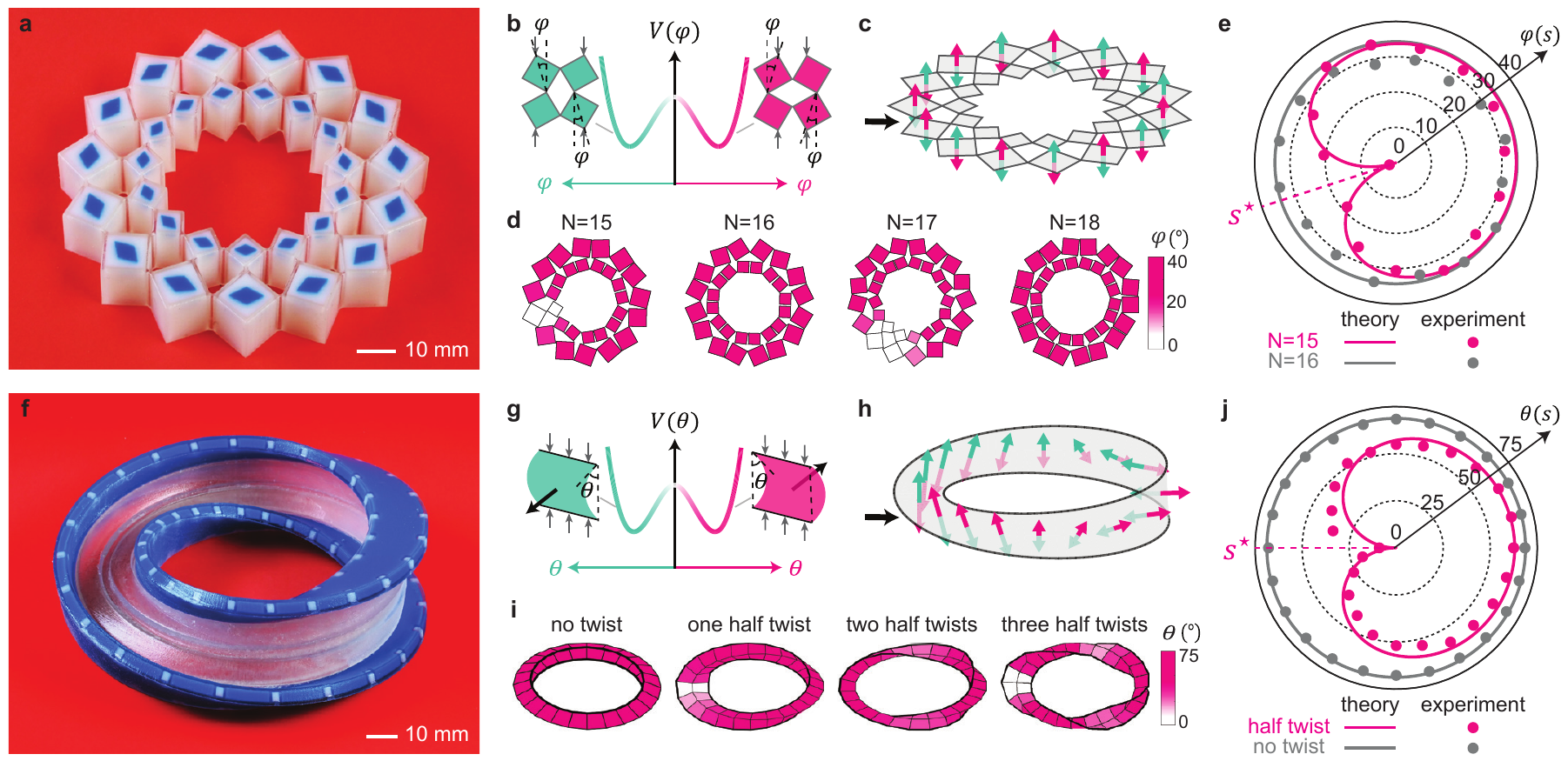}
    \caption{{\bf Non-orientable order in frustrated metamaterials and flexible M\"obius bands.} %\xg{version1} 
        {\bf (a)} A 3D-printed  metaring, which consists of 15 pairs of lozenges. 
    {\bf (b)} Plot of the double-well potential  $V(\varphi)$ that  distinguishes the two directions of rotation (green and pink). In practice, the potential is  approximated by a quartic function $V(\varphi)=K\xi^{-2}\left(\varphi^2-\varphi_0^2\right)^2$. 
    {\bf (c)} The arrows indicate the two alternative conventions defined in {\bf (b)} for the staggered-rotation vector of an odd metaring. No convention can be consistently defined over the whole ring. 
    {\bf(d)}  Experimental measurements of the staggered rotation angle $\varphi$ of metarings including even and odd number of units. Under homogeneous pressure, the deformations of odd metarings vanishes at an  unspecified node. Translation symmetry is spontaneously broken. 
    {\bf(e)} 
    Polar plot of $\varphi$ for an odd metaring with 15 pairs of lozenges (pink disks) and an even metaring with 16 pairs of lozenges (gray disks). 
    The solid lines are fits to the continuum theory detailed in SI. See also Supplementary Video 2.
    {\bf (f)} A 3D printed M\"obius band having a cross section in the shape of an I letter. The blue part is rigid in order to apply pressure while the white part is soft and can buckle. 
    {\bf (g)} Plot of the double-well potential  $V(\theta)$ that  distinguishes the two buckling directions of  flexible bands (shown in green and pink colour). 
    {\bf (h)} The pink and green arrows represent the normal to the M\"obius band corresponding to the two sign conventions defined in {\bf (g)}. No smooth normal vector field can be defined over the whole M\"obius band. 
    {\bf (i)}  Experimental measurements of the bending angle $\theta$ of  twisted  bands compressed by a homogeneous pressure. Under homogeneous pressure, the bending deformations of bands featuring an odd number of twists vanishes at an unspecified point. Translation symmetry is spontaneously broken.
     {\bf (j)} 
    Polar plot of the deformations of a M\"obius band (pink) and of a cylinder (grey). 
    The solid lines are fits to the continuum theory, see SI. See also Supplementary Video 3.
    }
    \label{fig:1}
\end{figure*}

The simplest instance of non-orientable order occurs in frustrated 1D antiferromagnets. To investigate their ground states experimentally, we use the mechanical metamaterial shown in Fig.~\ref{fig:1}a.
This structure consists of a closed metaring of $N$ pairs of coupled rotating lozenges, which are designed to promote counter rotation in response to external stresses~\cite{Bertoldi2017,Bertoldi_PRL2014,Coulais_NatPhys2017,Celli_softmatter2018,Choi2019,Zhang_Nature2019,Martin_NatPhys2020,Deng_PNAS2020,Martin_NatPhys2020,Yair_ice2021}. 
Under a sufficiently large axial compression, each pair of lozenges undergoes a structural instability (Fig.~\ref{fig:1}b). 
This transition is naturally characterised by a $\mathbb{Z}_2$-breaking order parameter: the staggered rotation angle of the lozenges $\varphi$ (Fig.~\ref{fig:1}b and c). 
In the lowest energy state, it can take two distinct values that correspond to the local minima of a double-well energy potential $V(\varphi)$ (Fig.~\ref{fig:1}b and  SI).

%%%%%%%

We now address the consequences of global frustration on the configurations accessible to the order parameter  $\varphi$ in the ordered---deformed---state.
To see this, we  apply a homogeneous load to the metaring by placing it in a vacuum bag, see Supplementary Video 2. A local direction of rotation is spontaneously picked up, the local $\mathbb Z_2$ symmetry is broken (Fig.~\ref{fig:1}d).
When the number  of pairs of lozenges $N$ is even, the order parameter $\varphi$ is homogeneous, and antiferromagnetic order extends over the whole structure (Fig.~\ref{fig:1}d and e).
Conversely, when $N$ is odd, the translation symmetry along the curvilinear coordinate $s$ is spontaneously broken, the order parameter $\varphi(s)$ is heterogeneous and vanishes at an unspecified point $s^\star$ (Fig.~\ref{fig:1}d and e). 
The resulting extensive degeneracy of the ground states is remarkable as the instability only breaks a discrete symmetry.  
To elucidate this counter-intuitive observation, we note that, when $N$ is odd there is an obstruction to define $\varphi(s)$ in a coherent fashion over the entire sample (Fig.~\ref{fig:1}c). 
Further insights can be gained from the continuum limit. The minimal elastic energy density derived in the SI 
\begin{equation}
    \mathcal E[\varphi(s)]= \frac{K}{2}(\partial_s\varphi)^2+V(\varphi),
    \label{eq:elastic}
\end{equation}
where $K$ is the elastic stiffness, describes well our measurements (Fig~\ref{fig:1}e and SI). 

In this continuum limit, the order parameter field belongs to a real line bundle (Fig.~\ref{fig:0}b), and the obstruction to define non-vanishing staggered-rotations over the whole metaring  implies that  this bundle is non-orientable, see~\cite{Hatcher2002} and SI.
This topological structure is encoded in the first Stiefel-Whitney index $w_1$, and defines two topologically distinct classes of order. 
When $N$ is even, the order parameter bundle is topologically trivial, $w_1=0$, and there is no obstruction to homogeneous ordering (Fig.~\ref{fig:1}d).
By contrast, when $N$ is odd, the order parameter bundle  is  non-trivial, $w_1=1$, and therefore  non-orientable~\cite{Hatcher2002}. 
Regardless of the parity of $N$, the antiferromagnetic metamaterial locally breaks a $\mathbb Z_2$ symmetry. 
Metarings with $w_1=0$ thus possess the two conventional ground states. On the other hand, metarings with $w_1=1$ possess an extensive number of ground states---enriched by non-orientable topology. 

This essential result immediately  suggests an alternative strategy to engineer non-orientable order using flexible M\"obius bands (Fig.~\ref{fig:1}f). 
The Euler buckling of a flat flexible band illustrated in Fig.~\ref{fig:1}g, provides a canonical example of a spontaneous $\mathbb Z_2$ symmetry breaking~\cite{Zdenek2010}.
Under the action of a sufficiently large axial load, it can  bend with equal probability along one direction or the other. 
By contrast,  when the band is twisted into a M\"obius strip, there is an intrinsic ambiguity to define a consistent orientation---hence a bending direction---over the whole band (Fig.~\ref{fig:1}h).
We experimentally and theoretically demonstrate that whenever the twisted bands are non-orientable, the bending amplitude belongs to a non-orientable bundle and a zero deformation node $s^{\star}$ of continuous degeneracy emerges and thus gives rises to non-orientable order (Fig.~\ref{fig:1}i and j, Supplementary Video 3 and SI for theory).

In stark contrast with the antiferromagnetic metamaterial, this robust property is inherited from the non-orientable shape of the M\"obius band itself. 
This requirement limits potential applications to (quasi)one-dimensional structures:
extending non-orientable mechanics to higher-dimensional bodies would require engineering surfaces like the Klein bottle and real projective plane, which cannot be realised  in practice. 
To circumvent this fundamental limitation, we  show below how to  engineer non-orientable order in 2D orientable frustrated metamaterials.

%%%%
\begin{figure*}[t!]
    \centering
    \includegraphics[width=0.8\textwidth]{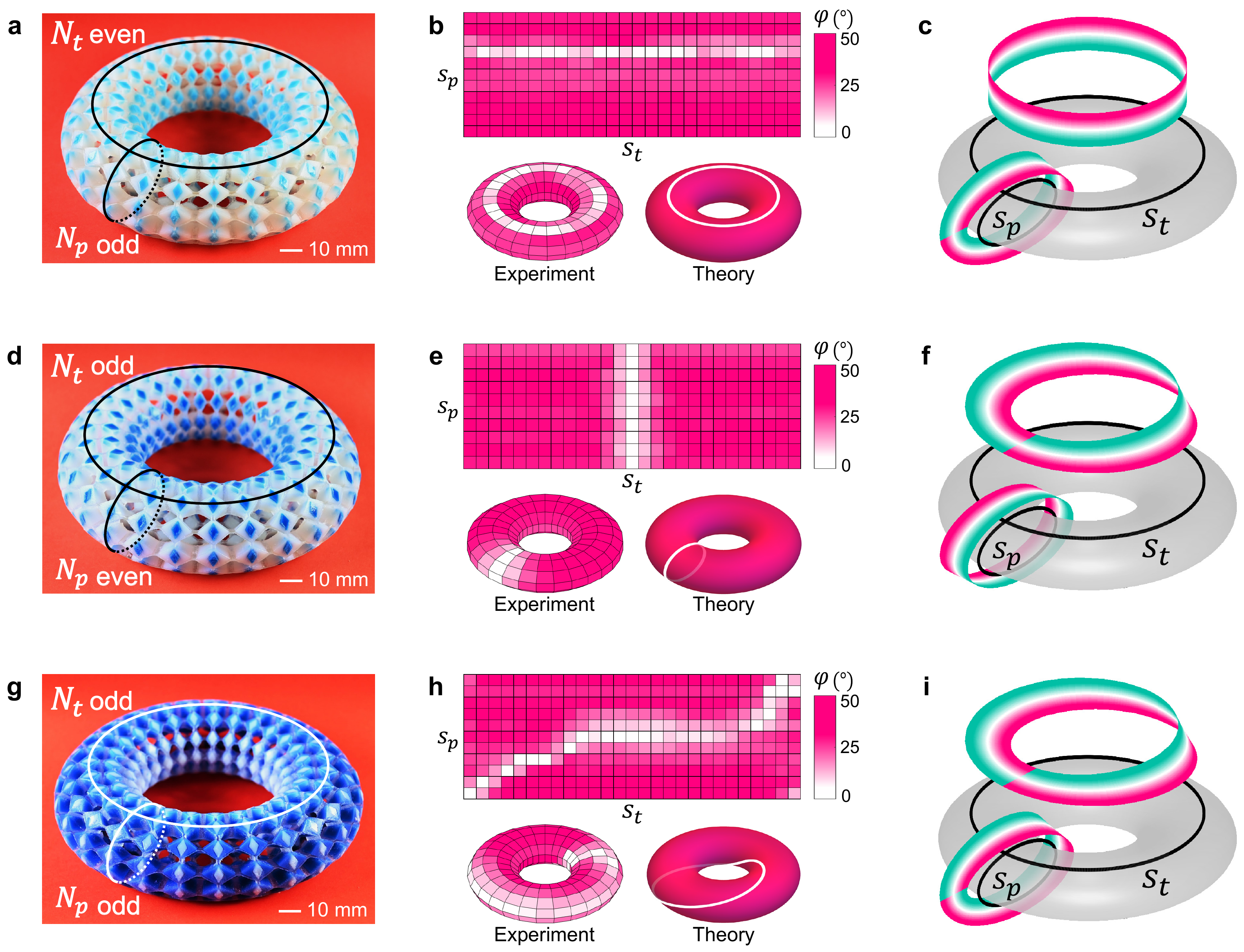}
    \caption{{\bf Non-orientable mechanics of toroidal metamaterials.} We vary the parity of the number of lozenges in the toroidal ($N_t$) and poloidal ($N_p$) directions: {\bf (a-c)} $N_t =26$ and $N_p =11$; {\bf (d-f)} $N_t =27$ and $N_p =10$; {\bf (g-i)} $N_t =27$ and $N_p =11$. {\bf (a, d, g)} 3D printed metatori. {\bf (b, e, h)} Experimental measurements of the staggered rotation field $\varphi(s_t,s_p)$  under homogeneous pressure, and theoretical prediction of the shape of the loop of zero deformation $\mathcal L$ (solid white line), see also Supplementary Video 4. {\bf (c, f, i)} Non-orientable bundles of the staggered-rotation fields. The gray tori represent the base space. The coloured strips illustrate the twist of the bundles along the toroidal and poloidal directions.}
    \label{fig:2}
\end{figure*}
%%%

To generalise non-orientable order to higher dimension, we construct toroidal metamaterials and tune the parity of the number of lozenges $N_{\rm p}$ and $N_{\rm t}$ along the poloidal and toroidal directions $s_{\rm p}$ and $s_{\rm t}$ to globally frustrate the $\mathbb Z_2$ order of their ground states (Fig.~\ref{fig:2}a,d and g). 
As illustrated in Fig.~\ref{fig:2}b,e and h, when we place a metatorus in a vacuum bag to apply a uniform compressive load, we observe the emergence of a loop of zero deformations. 
When $N_{\rm p}$  and $N_{t}$ have different parities, the loop winds along the even direction. 
When both directions are odd the loop takes an unanticipated helix shape, which spontaneously breaks mirror symmetry along both directions.

To explain how non-orientability sets the topology of the zero-deformation loops, we first consider the situation where $N_{\rm p}$ is odd and $N_{\rm t}$ even. Non-orientable order frustrates the emergence of a uniform ground state along the poloidal direction $s_{\rm p}$. 
Therefore, for any $s_{\rm t}$, there must exist a point $s_{\rm p}^{\rm \star}(s_{\rm t})$ where  $\varphi$ vanishes. 
This set of points defines a loop $\mathcal L$ of zero deformations, which winds once along the toroidal direction.
As a result of global frustration, the 
 staggered-rotation bundle is non-orientable, as illustrated in Fig.~\ref{fig:2}c and detailed in SI. 
The same reasoning readily applies to the two other classes of tori (Fig.~\ref{fig:2}f and i):
the non-orientable order of the metatori protects them from homogeneous deformations via the emergence of non-contractible $\mathcal L$ loops that cut the tori along the frustrated directions at an arbitrary location (Fig.~\ref{fig:2}b, e and h). 
\begin{figure*}[t!]
    \centering
    \includegraphics[width=\textwidth]{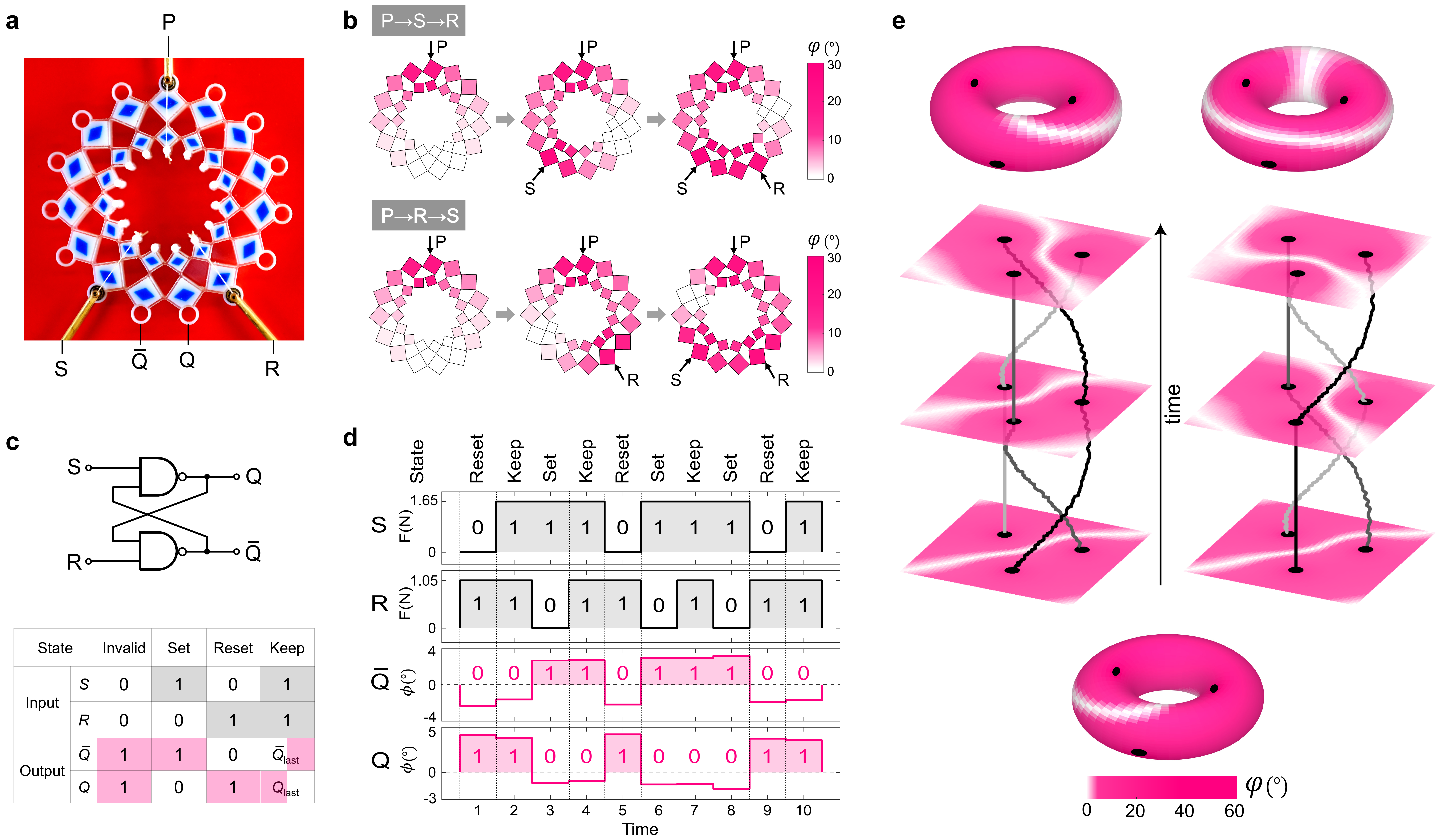}
    \caption{{\bf Non-abelian mechanics and mechanical memory.} 
    {\bf (a)} Picture of an odd metaring under point loads P, S and R (see SI for details). 
    {\bf (b)} Measurements of the staggered-rotation field $\varphi$ for two sequences of loading $\rm P\rightarrow \rm S\rightarrow \rm R$ (top) and $\rm P\rightarrow \rm R\rightarrow \rm S$ (bottom), see also \EDF~\ref{Ex_fig_non_abelian_curve} and Supplementary Video 5.
    {\bf (c)} Circuit diagram (top) and value table (bottom) of a Set-Reset latch.  
    {\bf (d)}  Correspondence between the input loads S and R (black) and output rotations (pink)  plotted as function of time for the metaring showed in {\bf (a)}, see also Supplementary Video 6.     %
    {\bf (e)} Numerical simulations of an odd-odd metatorus under two different sequences of identical point loads. The points of application of the loads are continuously moved to  braid their worldlines. While the initial and final load configurations are identical, the shape of the line of zero deformation explicitly depend on the braid form by the load history, see also Supplementary Video 7 and SI for details on the simulation protocol).
    }
    \label{fig:3}
\end{figure*}

The topology of the deformation bundle determines the winding direction of the zero-deformation loops, but their shape and response to perturbations are set by the specific form of the elastic energy. 
We  use again a minimal model for the elastic response of the metatori by generalising  Eq.~\eqref{eq:elastic} to two dimensions, see SI for details. 
In the vicinity of $\mathcal L$, the amplitude of the staggered deformations is small over a length $\xi$.  
We can then approximate the total elastic energy by  $E= \gamma\int \!{\rm d\sigma}$, where $\sigma$ is the curvilinear coordinate along $\mathcal L$. 
This is the energy of an elastic string having a finite line tension $\gamma\sim K\varphi_0^2(1+2\varphi_0^2)/(2\xi)$.  
 $\mathcal L$ is therefore the loop of minimal length satisfying the winding constraints imposed by non-orientability, in  agreement with our experimental findings reported in Fig~\ref{fig:2}b, e and h.

We have shown how frustrated metamaterials generically achieve non-orientable order beyond one dimension.
We now show that the topologically protected zero-deformation nodes and lines realise the elementary units of robust mechanical memory based on non-Abelian response.
Storing, reading and erasing mechanical information require the deformations to depend on the history of the loading sequence ~\cite{Treml_PNAS2018,Nagel_review2019,Andres_dome2020,Fruchart_Nature2020,Jules_arxiv2021,Yair_ice2021,Bense2021,Chen_Nature2021}.
One strategy consists in applying multiple loads to a material having a non-Abelian response, deformations that depends on the sequential order of the loads. 
To achieve this property, we  apply point loads to odd metarings (Fig.~\ref{fig:3}a). 
Applying a first load P results in a zero node at the diametrically opposite location (Fig.~\ref{fig:3}b). 
 Applying two subsequent loads S and R steers the zero node counterclockwise when S is applied first, or clockwise when R is applied first (Fig.~\ref{fig:3}b). 
 In other words, the state of deformations cannot be inferred from the sole knowledge of the load positions, but depends on their sequential order. 
This basic example of non-Abelian response stems from non-orientable order, which allows the multiplicity of the mechanical equilibria when more than one point load is applied, see SI.
 
To realise the write, read and erase operations, we demonstrate a one-bit digital memory akin to a Set-Reset latch~\cite{Horowitz_book1989} exemplified in Fig.~\ref{fig:3}c.
%SR latch sketch and table
We define two loads (S and R) and two rotation angles ($\overline{\rm Q}$ and Q) as input and output signals respectively (Fig.~\ref{fig:3}a). The sequential loading steps and measurements shown in Fig~\ref{fig:3}d realise all the Set-Reset latch operations shown in Fig~\ref{fig:3}c. 
We also stress that the read, write and erase operations can be performed sequentially without mechanical resetting~\cite{Treml_PNAS2018}.

In two dimensions, the zero-deformation loops protected by non-orientable order store information about the braiding of the point-load trajectories. %
We demonstrate this mechanical property in Fig.~\ref{fig:3}e, where we show how to tailor the morphology of $\mathcal L$ loops. Identical sets of local loads result in dramatically different deformation states that reflect the full braiding history of their trajectories. 
We therefore expect non-orientable order to offer an avenue to perform computational tasks based on source braiding alternative to the holonomic computing paradigm~\cite{Zanardi1999,alicea2011non,Fruchart_Nature2020}.

\emph{Acknowledgments.} 
 We thank Jasper van Wezel, Martin van Hecke, Anne Meeussen, Li Ma and Yair Shokef for insightful discussions and suggestions, Daan Giesen for technical assistance and Am\'elie Chardac, Camille Jorge and Romane Braun for help with the active matter and packing experiments. X.G. acknowledgdes financial support from the China Scholarship Council. D.B. and D.C. acknowledge support from IDEXLYON ToRe and ANR WTF grants. C.C. acknowledges funding from the European Research Council via the Grant ERC-StG-Coulais-852587-Extr3Me. 

All the codes and data supporting this study are available on the public repository  \url{https://doi.org/10.5281/zenodo.5730508}.

\setcounter{equation}{0}
\renewcommand{\theequation}{A\arabic{equation}}%
\setcounter{figure}{0}
\renewcommand{\thefigure}{A\arabic{figure}}%
\renewcommand{\figurename}{\EDF}%

% \newpage
\begin{appendix}

\begin{figure*}[t]
    \centering
    \includegraphics[width=0.7\textwidth]{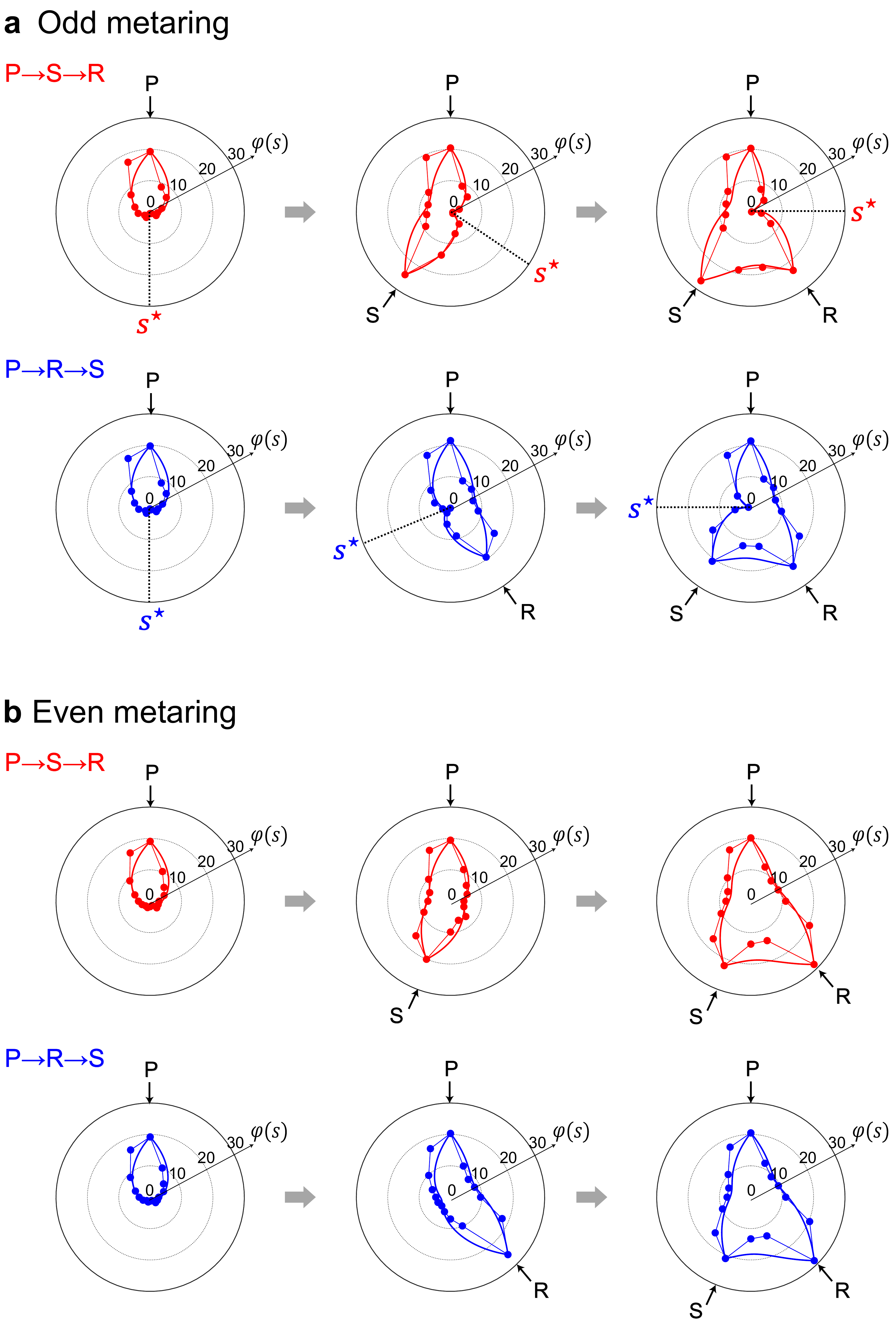}
    \caption{\textbf{Polar plots of the staggered rotation angle $\bm\varphi$ of an odd (even) metaring under different loading sequences $\bm{P\rightarrow R\rightarrow S}$ and $\bm{P\rightarrow S
\rightarrow R}$. } {\textbf{(a)} Under two sequences of loading, an odd metaring (15 pairs of squares) exhibits different responses. The zero node is marked as $s^{\star}$. \textbf{(b)} Under two sequences of loading, an odd metaring (15 pairs of squares) displays the same response. Scatter plots show experimental measurement. The solid lines are fits to the continuum theory, see SI.}}
    \label{Ex_fig_non_abelian_curve}
\end{figure*}

\end{appendix}

% SUPPLEMENTARY INFORMATION

\clearpage 
\onecolumngrid 
\begin{center}
 \Large
{\bf Supplementary Information }
\normalsize
\end{center}

\newcommand{\RN}[1]{\textup{\uppercase\expandafter{\romannumeral#1}}}
\newcommand{\beq}{\begin{equation}}
\newcommand{\eeq}{\end{equation}}
\newcommand{\com}[1]{\textcolor{red}{[#1]}}
\newcommand{\mtf}{of the Main Text}
\definecolor{macouleur}{RGB}{105,150,150}

%Figure SX
\counterwithout{equation}{section}
\setcounter{equation}{0}
\renewcommand{\theequation}{S\arabic{equation}}%
\setcounter{figure}{0}
\renewcommand{\figurename}{{Fig.}}
\renewcommand{\thefigure}{{S\arabic{figure}}}

\section{Supplementary Videos}
Supplementary Video 1: Active liquid shows global frustration.\par
Supplementary Video 2: An odd metaring and an even metaring are compressed homogeneously by a vacuum bag.\par
Supplementary Video 3: A M\"obius band and a cylinder are compressed homogeneously by a vacuum bag.\par
Supplementary Video 4: Tori with different parities of the number of lozenges along toroidal and poloidal directions are compressed homogeneously by a vacuum bag. \par
Supplementary Video 5: Under two different loading sequences, an odd metaring shows non-Abelian response.\par
Supplementary Video 6: A mechanical sequential logic gate realises all the Set-Reset latch operations. \par
Supplementary Video 7: Numerical
simulations of an odd-odd metatorus under different sequences of identical point loads.

\section{Non-orientable bending mechanics of M\"obius strips}
\subsection{A M\"obius strip is a non-orientable surface}
\label{sec:mobius}
\begin{figure}[h!]
\includegraphics[width=1\columnwidth,angle=0]{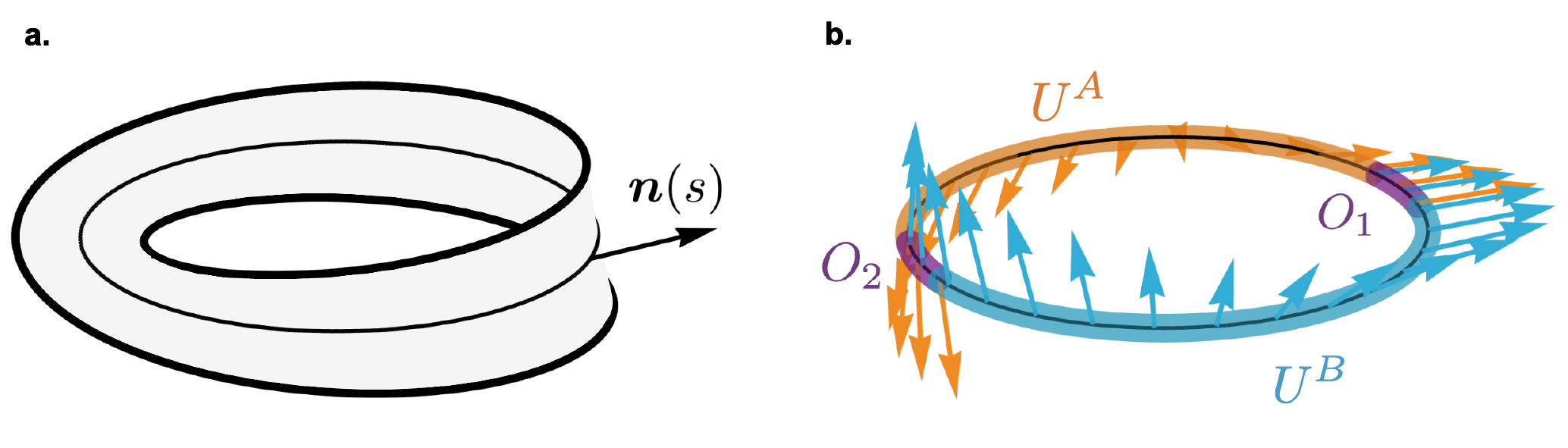}
\caption{{\bf M\"obius strips are non-orientable}. {\bf a.} Ruled surface having the shape of a M\"obius strip, and definition of the normal vector $\bm n(s)$. {\bf b.} 
It is impossible to define a consistent choice of a continuous normal-vector field on a M\"obius strip. 
Considering two open sets covering the base circle $S_1$, the product of the transition functions over the two open overlaps $O_1$ and $O_2$ is equal to $-1$.}
\label{Fig:Mobius}
\end{figure}

Fig.~\ref{Fig:Mobius}a shows a M\"obius strip defined as a ruled surface of constant width $w$. We note $s$ the curvilinear coordinate along the circular centerline of unit length, and $\bm n(s)$ the unit vector normal  to the surface at $s$.
The M\"obius strip provides a prototypical example of  a non-orientable manifold: it is impossible to coherently define  a smooth normal vector field  over the whole strip.  
At best, we can define smooth unit normal vectors $n_{A}$ and $n_B$ over two overlapping open intervals $U^A$ and $U^{B}$, as illustrated  in Fig.~\ref{Fig:Mobius}b.  
In each of the two overlap regions  $O_{1}$ and $O_{2}$, we can define a transition function $\eta_{1,2}$ which relates the two orientation conventions $\bm n^A$ and $\bm n^B$:  $\bm n^A=\eta^{AB}_{1}\bm n^B$ in $O_{1}$ and $\bm n^A=\eta^{AB}_{2}\bm n^B$ in $O_{2}$. Non-orientability translates in the relation
~\cite{BartoloCarpentierPRX}:
\begin{equation}
\eta^{AB}_{1}\eta^{AB}_{2}=-1,
\label{Eq:eta}
\end{equation}
which prevents the definition of a smooth unit-vector field $\bm n(s)$ over $U^A\cup U^B$~\cite{Gramain,Hatcher2002}. 
We explain below that the non-orientability of soft M\"obius strip's results in a non-trivial topology of their bending deformations bundle.
 %%%%%%%%%%%%%%%%
%%%%%%%%%%%%%%%%
\subsection{The bending-deformation bundle of a M\"obius strip is non-orientable}
\begin{figure}
\includegraphics[width=\columnwidth,angle=0]{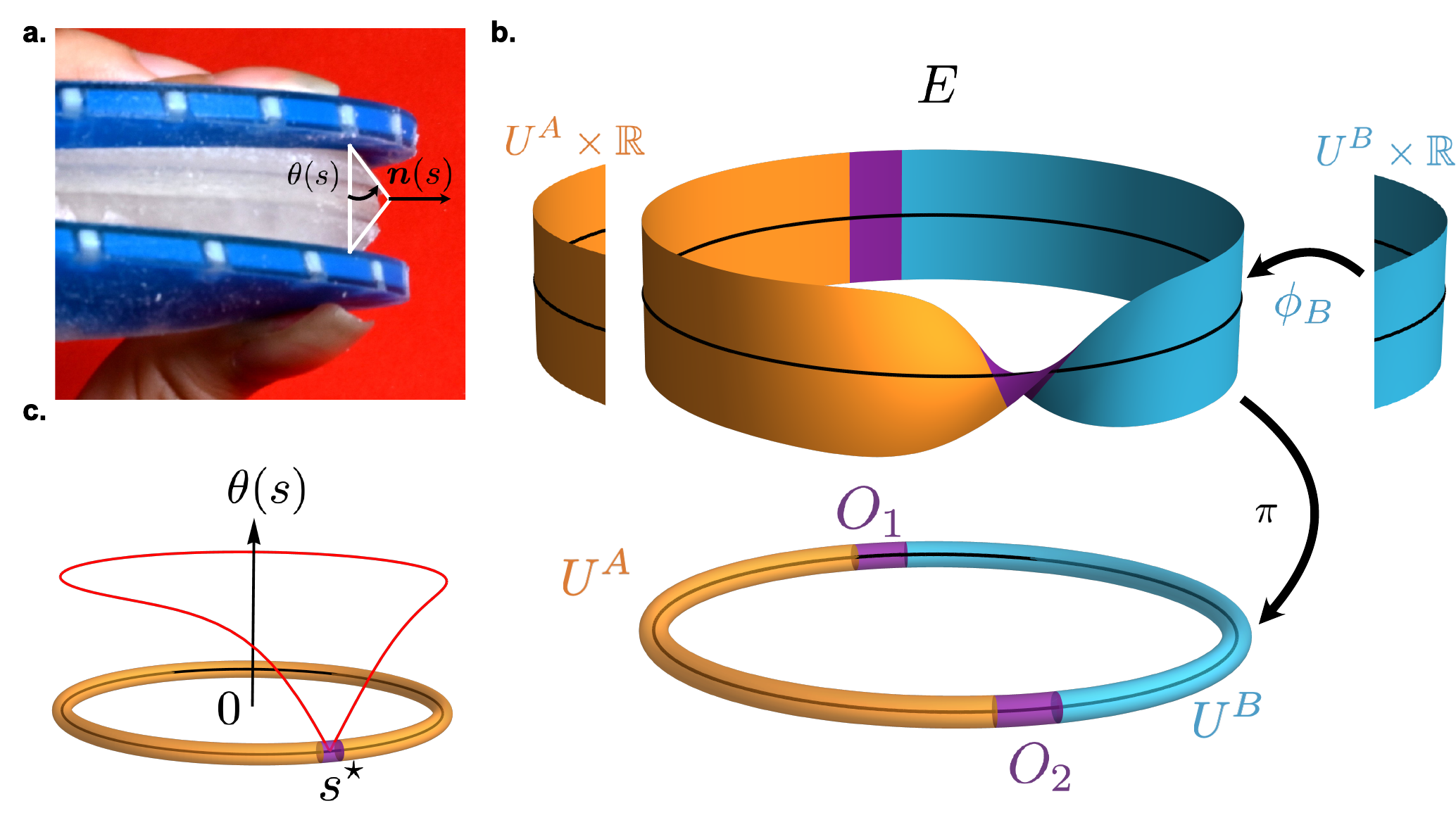}
\caption{{\bf Bending-deformation bundle of an elastic M\"obius strip}. {\bf a.} Picture of a buckled M\"obius strip described in the Methods.
In our simplified picture, the bending elasticity is captured by a single bending angle $\theta(s)$, the resulting local displacement vector is $\delta \bm R^{\star}(s)=\sin\theta(s) \mathbf n(s)$. 
{\bf b.} Sketch of the line bundle structure.
The total space $E$ is locally isomorphic to a cylinder ($\mathbb R\times S_1$) but not globally. 
The two sections $\theta_{\rm A}$ and $\theta_{\rm B}$ are defined on two overlapping open intervals $U^A$ and $U^B$ of the unit circle $S_1$. 
{\bf c.} Maximally large trivial section of the  deformation bundle. It is defined over $U^{\rm A}=S_1\backslash\{s^\star\}$, the non-orientability of the bundle structure implies that the deformation vanishes at $s^\star$.
}
\label{Fig:MobiusBundle}
\end{figure}

To single out the impact of non-orientability on  buckling deformations,  we use a  low-energy description of the bending deformations. 
More specifically, we consider a single bending angle $\theta(s)$ associated to a displacement $\delta \bm R^{\star}(s)/w=\frac1 2 \sin\theta(s) \bm n(s)\sim\frac 1 2 \theta(s)$ along the local normal vector, see Fig.~\ref{Fig:MobiusBundle}a. The  bending deformations fields $\theta(s)$ define a real line bundle over $S_1$. We now show that its topology is non trivial. 

We first note that, as $\bm n(s)$ must be defined separately on the two open sets $U^A$ and $U^{B}$, we also have to assign two consistent sign conventions for the bending angle ($\theta_{\rm A}$ and $\theta_{\rm B}$).
Conversely, the physical observable $\delta \bm R^{\star}(s)$ is  a displacement vector in our  3D Euclidian space, which does not depend on any local representation of the M\"obius strip orientation. 
Therefore, when the orientation convention  changes, the deformations must  obey the same $\mathbb Z_2$ gauge transformation rule~\cite{BartoloCarpentierPRX}:
\begin{equation}
\label{Eq:Z2}
    \left \{
    \begin{array}{ccc}
        \bm n_{\rm A/B}&\to& -\bm n_{\rm A/B} \\
        \theta_{\rm A/B}&\to &-\theta_{\rm A/B}
    \end{array}
    \right.
\end{equation}
In all that follows $X_{A/B}$ stands for $X_A$ or $X_B$.
We can now  describe the strip elasticity in term as a non-trivial line bundle, sketched in  Fig.~\ref{Fig:MobiusBundle}b, see~\cite{Gramain,Hatcher2002} for more mathematical details. 
%%%%%%%%%%%%%%%%
%%%%%%%%%%%%%%%%
The local deformation $\theta(s)$ belongs to a real fiber $\mathbb R_s$ at each point $s$ of the base manifold $S_1$.  
By definition, the  total space $E$ of the deformations defines a fiber  bundle. The scalar deformation fields  ${\theta_{A/B}(s)}$ define two local sections of the deformation bundle.
In an open neighborhood  $U_s$ of $s$, the deformations $\theta(s)$ is defined by a local choice of orientation. In more mathematical terms, this defines a local trivialization $\pi^{-1}(U_s)=\Phi_s(U_s\times \mathbb R)$, where $\Phi_s$ is a smooth mapping, and where $\pi$ is the local projection $E  \xrightarrow{\pi } S_1$.
Finally, the local choice of the orientation of the M\"obius strip, discussed in Section~\ref{sec:mobius},  naturally associates a $\mathbb Z_2$ structure group   (Eq.~\eqref{Eq:Z2}) to the bundle definition, see also~\cite{BartoloCarpentierPRX}. 

We can see that the deformation  bundle is non-trivial by using the two bundle charts $(U^A,\Phi^A)$ and $(U^B,\Phi^B)$ sketched in Fig.~\ref{Fig:MobiusBundle}b. In the open overlap  $O_1$, we can always choose the same orientation for the two charts, i.e $(\Phi^A)^{-1}\circ\Phi^B (s,\theta)=(s,\theta)$ for all $s\in O_1$. 
However, given this choice,  Eq.~\eqref{Eq:eta}, and the independence of the real-space displacements $\delta \bm R^{\star}(s)$ on the  local orientation of the strip, imply that $(\Phi^B)^{-1}\circ\Phi^A(s,\theta)=(s,-\theta)$. 
The transition function $\Phi^B\circ(\Phi^A)^{-1}$ relating the two bundle charts in $O_2$ must have a negative Jacobian. 
This result defines a non-orientable line bundle: the bending deformations of a soft Möbius strip is non-orientable~\cite{Gramain,Hatcher2002}. 
An essential property of real line bundles is that non-orientability is equivalent to topological non-triviality. When a bundle is non-trivial, it is impossible to define a smooth non-vanishing section $\theta(s)$  over the whole base space $S_1$. Regardless of the specifics of the elastic energy,  non-orientability  requires  continuous bending deformations $\theta(s)$ to vanish at least at one point $s^\star$ along the strip. 
Simply put, M\"obius strips are topologically protected against homogeneous buckling. 
 
 We close this section with three comments. 
 Firstly, in the Main Text, and in all that follows, we avoid referring to the definition of two separate intervals by defining $U^A$ as the maximally large open set over which the line bundle can be trivialized.
 Given the location of $s^\star$, we define {\it a posteriori} 
 $U^{\rm A}=S_1 \backslash \lbrace s^\star\rbrace$. We then drop the $A$ index and $\theta_{\rm A}(s)\equiv\theta(s)$. 
Given this definition of the bending deformations, $\theta(s)$ is continuous  over the whole strip and vanishes at $s^\star$, Fig~\ref{Fig:MobiusBundle}c. 
Secondly, the  definition of $\theta (s)$ does not require introducing a double covering of the M\"obius strip~\cite{Goldstein2015}. 
The double-covering formulation is not in contradiction with the one used in this work. They represent two line-bundle representations of topologically constrained scalar fields  defined over $S_1$. 
Finally, our representation of the deformations of non-orientable manifolds are  based on single-valued continuous fields. Therefore, it
allows us to use a standard elastic elastic energy: 
 \begin{equation}
 \label{eq:Emobius}
     \mathcal E=\frac 1 2\int_{S_1\backslash \{s^\star\}} \!\!\!\!\!\!\!\!\!\!\!\!\!\! (\partial_s \theta)^2+2V(\theta)\,{\rm d}s,
 \end{equation}
 where $V(\theta)$ is a potential parametrized by the magnitude of the axial load $F$. $V(\theta)$ is a symmetric bistable potential when $F$ exceeds the Euler buckling threshold, see Fig. 1. The minimization of Eq.~\eqref{eq:Emobius} in the specific case of a quartic potential was discussed in~\cite{BartoloCarpentierPRX}. The comparison to our experimental measurements  (Fig. 1d in the Main Text) demonstrates that the minimal model introduced in this section provides an excellent proxy of the full thin-sheet elasticity problem~\cite{Audoly}.

\section{Non-orientable mechanics of anti-ferromagnetic mechanical metamaterials\label{sec.BSModel}}

\subsection{Bending-Shearing model}
The mechanical description of the metamaterials showed in Figs. 1 and 2 of the Main Text, is based on the bending-shearing model of Ref.~\cite{Coulais_NatPhys2017}. 
This minimal model describes the deformations of a lattice of of rigid lozenges connected by elastic hinges, see Fig.~\ref{fig.SI1}a. 
In this picture, the lozenges are squares of size $L$, their centers are separated by a distance $a$ and the hinges' length is $\ell$.
The predominant deformations  are the bending and shearing modes of the elastic hinges sketched in Fig.~\ref{fig.SI1}b. 
In the limit of small deformations, we neglect the translation of the lozenge centers, and focus on their rotation. The Bending-Shearing model therefore  consists in a  competition between co-rotation of neighboring lozenges promoted by bending, and counter-rotation  promoted by shearing, Fig.~\ref{fig.SI1}b.

\subsection{Continuum mechanics of an open meta-chain}
%%%%
\begin{figure}
\begin{center}
	\includegraphics[width=1\columnwidth]{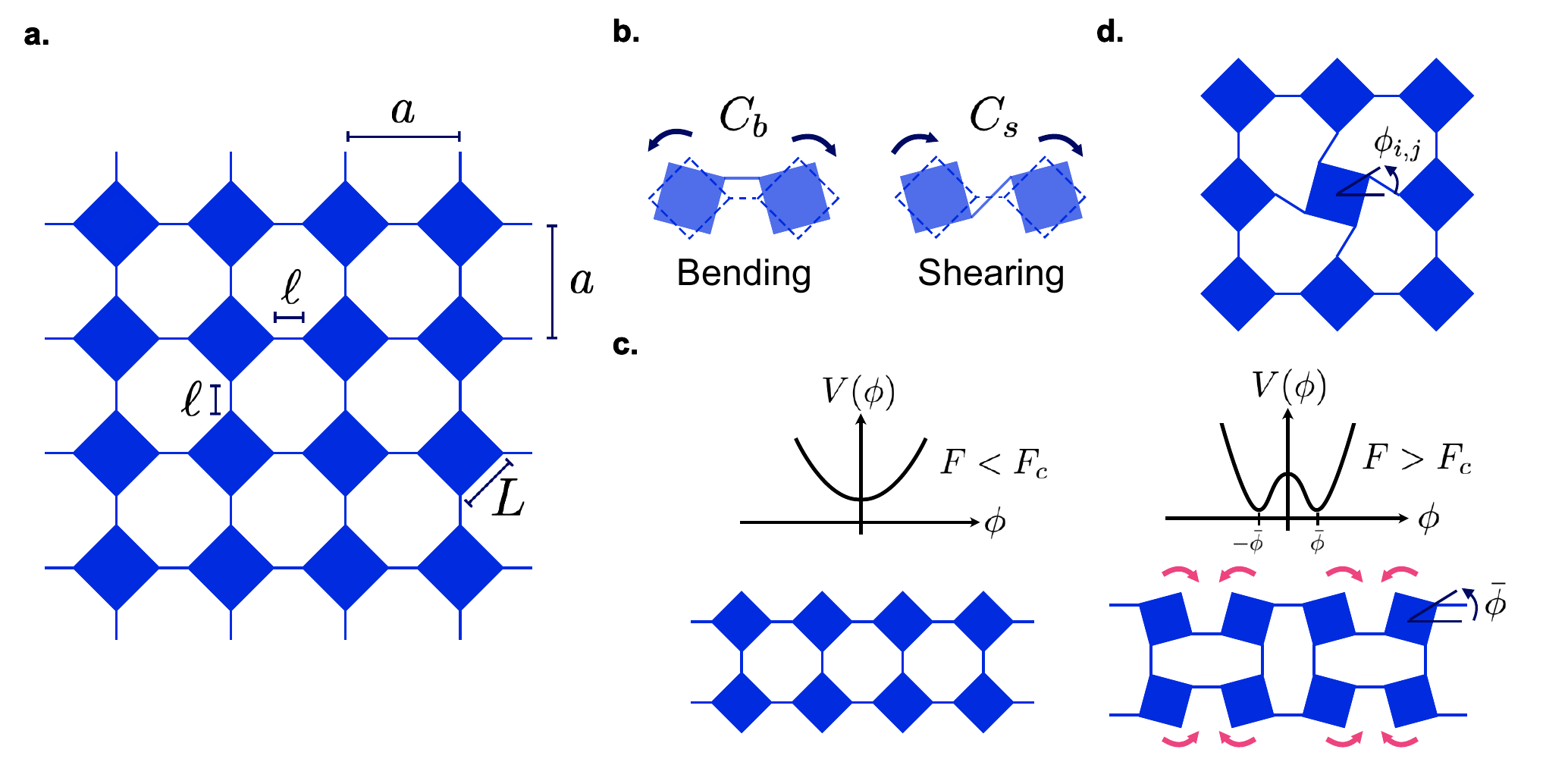}
\end{center}
\caption{{\bf Bending-Shearing model. model}
{\bf a.} A square lattice of rigid squares connected by soft hinges. We note: $a$ the lattice spacing, $\ell$ is the hinge length, and $L$ the square size.
{\bf b.} Bending and shearing deformations of the filaments promote respectively the co-rotation and counter-rotation of adjacent squares.
{\bf c}
Each square is associated to a single degree of freedom $\phi$ corresponding to its rotation with respect to its equilibrium position. 
{\bf d.} Total mechanical energy of the metamaterial under the action of a homogeneous compression load $F$. When $F<F_{\rm c}$ the lattice is  undeformed. When $F>F_{\rm c}$, the lattice buckles in plane, and a global anti-ferromagnetic order emerges. 
}
\label{fig.SI1}
\end{figure}
%%%
 To gain some intuition, we  consider first the simple case of an open meta-chain made of two rows  of $N$ squares. 
 Within a harmonic approximation, and assuming mirror symmetry between the two rows, the elastic energy reduces to
\beq
E=2\sum_{i=1}^{N-1}\left[\frac{C_b}{2 a}(\phi_{i+1}-\phi_i)^2+\frac{C_s}{2 a}\left(\phi_{i+1}+\phi_i\right)^2\right]+\sum_{i=1}^N \frac{C_b}{2 a}(2\phi_i)^2,
\label{eq.EnergyMeta1D}
\eeq
where the rotation $\phi$ is defined in Fig.~\ref{fig.SI1}.
The first sum reflects the competition between co-rotation  and counter-rotation. The two coupling constants are given by $C_b=k_b$ and $C_s=\frac{k_s}{4}(1+\sqrt{2}L/\ell)$, where  $k_b$ and $k_s$ are the bending and shearing stiffness of the hinges, see~\cite{Coulais_NatPhys2017}. 
The second sum in Eq.~\eqref{eq.EnergyMeta1D} arises from the couplings between the two symmetric rows, and hinders the rotation of each individual square.

The mechanics of the metamaterial has an obvious magnetic analogy. 
The first sum in Eq.~\eqref{eq.EnergyMeta1D} mirrors  the competition between ferromagnetic and anti-ferromagnetic interactions in a collection of $XY$ spins, within a spin wave-approximation. 
The second sum mirrors the role of  a homogeneous  magnetic field. 
All of our  experimental  results correspond to situations where $C_b\ll C_s$, we therefore limit our analysis to this case, and dub our mechanical systems   anti-ferromagnetic metamaterials.

In addition to the elastic energy, the mechanical work $W$ of an  axial load $F$ contributes to the total buckling energy $\mathcal E=E+W$ as 
\begin{equation}
W=F a\sum_i (\cos \phi_i-1),
\label{eq:work}
\end{equation}
When $F<F_{\rm c}$, with $F_{\rm c}=12C_{\rm b}/a^2$,  the total energy $\mathcal E$
has a single minimum which corresponds to a homogeneous state with $\phi_i=0$, Fig.~\ref{fig.SI1}d. But, when $F$ exceeds $F_{\rm c}$, the homogeneous rest state becomes unstable, the squares rotate rotate to form a staggered structure depicted Fig.~\ref{fig.SI1}d. The metamaterial undergoes an anti-ferromagnetic transition. 

To describe the low energy excitations of an open chain, we naturally  introduce the local anti-ferromagnetic order parameter $\varphi_i=\epsilon(-1)^i\phi_i$, where $\epsilon=\pm1$ is an arbitrary sign convention for the rotation direction. 
Close to the onset of the anti-ferromagnetic transition, we can expand  the cosine function up to quartic order in Eq.~\eqref{eq:work}, and write $\mathcal E$ in the canonical form:
\beq
\mathcal E=\sum_i\left[\frac{K}{2a}(\varphi_{i+1}-\varphi_i)^2+a V(\varphi_i)\right],
\label{eq:EAFdiscrete}
\eeq
with $K= 2(C_s-C_b)$ is the anti-ferromagnetic stiffness, and $V(\varphi)=K/\xi^2(\varphi^2-\varphi_0^2)^2$, with $\xi^2=24K/F$ and $\varphi_0^2=6(1-F_{\rm c}/F)$. 
In the long wave-length limit, we can therefore describe the anti-ferromagnetic transition in term of a canonical $\varphi^4$ model:
\begin{equation}
    \mathcal E=\int {\rm d}s\, \frac{K}{2}(\partial_s\varphi)^2+V(\varphi),
    \label{eq:phi4}
\end{equation}
where $s$ is the curvilinear coordinate along the chain and $\varphi(s)$ is a smooth  staggered rotation field.

\subsection{Non-orientable mechanics of closed meta-chains}
\label{Sec:MetaRings}
\begin{figure}
\includegraphics[width=\columnwidth,angle=0]{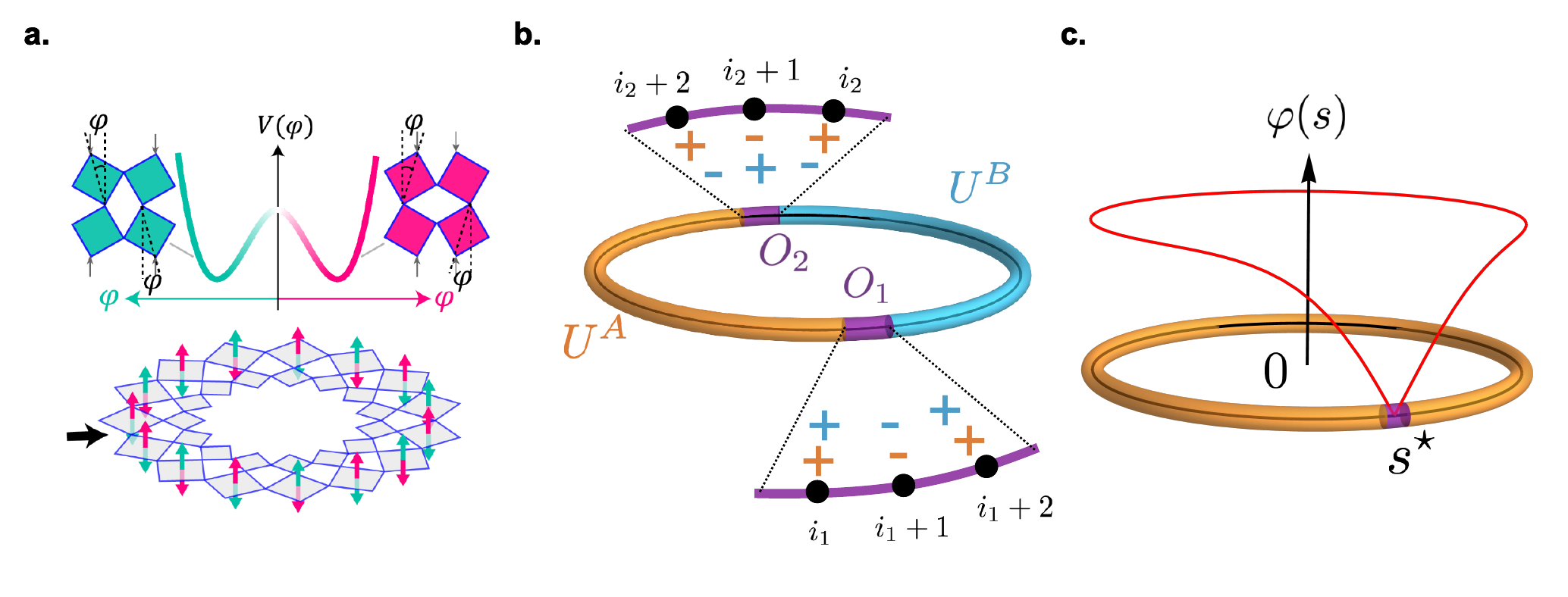}
\caption{{\bf frustration induced non-orientability of odd metarings}.
{\bf a.} In a closed metaring, anti-ferromagnetic order is frustrated when the number $N$ of lozenges is odd.
{\bf b}:  
We can cover the rig of lozenges by two overlapping sets where we  define two  orientations $\epsilon_A$ and $\epsilon_B$ of the staggered angles $\varphi_i^{A/B}$. %
The global frustration of the anti-ferromagnetic order implies that the transitions functions $\eta$ cannot be all equal to $1$ in the two overlap sets $O_1$ and $O_2$. 
{\bf c.} Continuum limit. 
The largest open set over which the elastic-deformation bundle can be trivialized is $S_1/\{s^{\star}\}$. 
At the position $s^{\star}$ the non-triviality of the deformation bundle imposes  $\varphi(s^{\star})=0$.}
\label{Fig:frustration}
\end{figure}

When defining the phonon elasticity of a periodic lattice, taking the continuum limit does not require caring about the parity of the number of atoms. 
In stark contrast, the long wavelength excitations of closed mechanical anti-ferromagnets crucially depends on the parity of $N$. 
As illustrated in Fig.~\ref{Fig:frustration}a,  a metaring with an odd number of lozenges  frustrates global anti-ferromagnetic order. 

We now show that  the geometrical frustration of mechanical deformations translates in the  non-orientability of their  associate bundle. 
As illustrated in Fig.~\ref{Fig:frustration}b,   we cover the closed chain by two overlapping open sets $U^A$ and $U^B$  corresponding to  two arbitrary orientation conventions $\epsilon^A=\pm1$ and $\epsilon^B=\pm1$ defining two staggered-rotations fields:
$\varphi_i^{A/B}=\epsilon^{A/B}(-1)^i\phi_i$. 
The local deformation  $\phi_i$ is defined unambiguously with respect to the vector normal to the planar metamaterial.
Therefore, the orientation and staggered rotation variables obey the same $\mathbb Z_2$ transformation rule as in Eq.~\eqref{Eq:Z2}:
\begin{equation}
\label{Eq:Z2bis}
    \left \{
    \begin{array}{ccc}
         \epsilon^{ A/B}&\to& -\epsilon^{ A/B}, \\
        \varphi^{ A/B}&\to &-\varphi^{A/B}. 
    \end{array}
    \right.
\end{equation}
We can always choose $\epsilon_A=\epsilon_B$ in the first overlapping region $O_1$ via e.g. a redefinition of the sign of $\varphi^B$.
However, in $O_2$ the transition function which relates the two  staggered rotations is determined by the parity of $N$.
A direct count of the number of sites  separating the two overlap regions yields: $\epsilon_A\varphi_i^A=(-1)^N\epsilon_B\varphi_i^B$, for $i\in O_2$.
When $N$ is even there is no obstruction to trivialize the deformation bundle into $S_1\times \mathbb R$, and 
the long wave-length elastic energy is given by Eq.~\eqref{eq:phi4}. By contrast, taking the continuum limit $N\to\infty$ and keeping $N$ odd,  defines a non-orientable deformation bundle. 
Following the exact same reasoning as in Section~\ref{sec:mobius}, we find that the staggered-rotation bundle $E  \xrightarrow{\pi } S_1$ is topologically identical to the bending-deformation bundle of a continuous M\"obius strip: the staggered-rotation bundle has an emergent non-orientable topology although the metamaterial itself is obviously orientable.
Non-orientable real line bundles are non-trivial, as a consequence we cannot define a smooth non-vanishing field $\varphi(s)$ over the entire chain when $N$ is odd, Fig.~\ref{Fig:frustration}.

Following the same reasoning as in Section~\ref{sec:mobius}, in  the Main Text we consider the largest possible trivialization of $E\xrightarrow{\pi}S_1$, see Fig.~\ref{Fig:frustration}c.
We  define $\varphi$  over the interval $U=S_1\backslash\left \{s^{\star}\right\}$, where $s^{\star}$ is the point where the deformation must vanish, Fig.~\ref{Fig:frustration}c:
\begin{equation}
    \varphi(s^\star)=0.
    \label{eq:sstar}
\end{equation}
We stress that the location of the zero-deformation node is  not  prescribed {\it a priori}, and
this additional (gauge) degree of freedom needs to be dealt with when  minimizing   the elastic energy, Eq.~\eqref{eq:phi4}, of odd metamaterials. 
%%%%%%%%%%%%%%%%
%
\subsection{Zero-deformation node on frustrated rings: a pedestrian demonstration}
\label{Sec.pedestrian}
%%%%%%%%%%%%%%%%
In this section we provide an alternative  demonstration of  Eqs.~\eqref{eq:phi4} and~\eqref{eq:sstar}. This more pedestrian approach starts with the lattice model defined by Eq.~\eqref{eq:EAFdiscrete}. As in the continuum approach, we define two sets  of lattice points $U^A$ and $U^B$ to cover the ring, Fig.~\ref{Fig:frustration}b. 
 We can write the anti-ferromagnetic part of the mechanical energy by keeping track of the regions over which the staggered deformations are defined:
\begin{align}
      E_{\rm AF}=\sum_{i\in{S_A}\backslash\{i_1\}} \frac{K}{2a} (\varphi_i^A-\varphi_{i+1}^A)^2+ aV(\varphi_i^A)+
      \sum_{i\in{S_B}\backslash\{i_2\}} \frac{K}{2a} (\varphi_i^B-\varphi_{i+1}^B)^2+ aV(\varphi_i^B).
      \label{Eq:EAB}
\end{align}
We can always find a couple of orientations $\{\epsilon^A,\epsilon^B\}$ so that   $\varphi^A(i_1)=\varphi^B(i_1)$, in the overlap $O_1$. 
However, this choice implies that $\varphi^A(i_2)=(-1)^N\varphi^B(i_2)$ in $O_2$. Given this observation, we can can drop the $A/B$ indices and express Eq.~\eqref{Eq:EAB} in term of a single deformation variable $\varphi_i$ as 
\begin{align}
      E_{\rm AF}=\sum_{{S_A\cup S_B}\backslash\{i_2\}} \frac{K}{2a} (\varphi_i-\varphi_{i+1})^2+ aV(\varphi_i)+
       \frac{K}{2a} (\varphi_{i_2}+\varphi_{{i_2}+1})^2.
      \label{Eq:EAB}
\end{align}

We then define $w_1=\frac 1 2 [1-(-1)^N]$ and  recast the above expression into 
\begin{align}
    E_{\rm AF}&=\sum_{i} \frac{K}{2a} (\varphi_i^A-\varphi_{i+1}^A)^2+aV(\varphi_i)+4w_1{\frac{K}{a}} \varphi^A_{i_{\rm G}}\varphi_{i_{\rm G+1}}^A,
\end{align}
where $a=1/N$ for a ring of unit length.
We can now take the continuum limit ($N\to\infty$) keeping the parity of $N$ constant, see Fig.~\ref{Fig:frustration}b.
The long wave-length description  of  anti-ferromagnetic chains then takes the compact form:
\begin{align}
\label{Eq:defsG}
    {\mathcal E}=\int\! \left[ \frac{K}{2}(\partial_s \varphi)^2 +V(\varphi)\right]\, {\rm d}s
    &+4K w_1\lim_{N\to\infty} N \varphi^2(s^{\star}),
\end{align}
where $s^\star=i_2/N$.
The second term of Eq.~\eqref{Eq:defsG} originates from the topology  of the deformation bundle. When $N$ is even, the deformations are orientable, $w_1=0$ and Eq.~\eqref{Eq:defsG} reduces to the  linear low-energy description of 1D elastic materials (Eq.~\eqref{eq:phi4}).
Conversely, when $N$ is odd, mechanics is non-orientable, $w_1=1$ and the last term of Eq.~\eqref{Eq:defsG} constrains any continuous deformation field $\varphi(s)$ to vanish at $s=s^{\star}$. 
From a more formal perspective $w_1$ corresponds to the first Stieffel-Whitney class of the elastic deformation bundle. It defines an obstruction to its trivialization~\cite{Hatcher2002}.

We showed in the previous section that the staggered deformations of frustrated odd chains are topologically identical to the bending modes of a soft M\"obius strip. This result is further confirmed by the structure of Eq.~\eqref{Eq:defsG}  which includes the same topological term as in the minimal  model of a M\"obius strip introduced in~\cite{BartoloCarpentierPRX}. 

\subsection{Buckling patterns of 1D metarings: comparison to experiments}
\begin{figure}
\begin{center}
	\includegraphics[width=1\columnwidth]{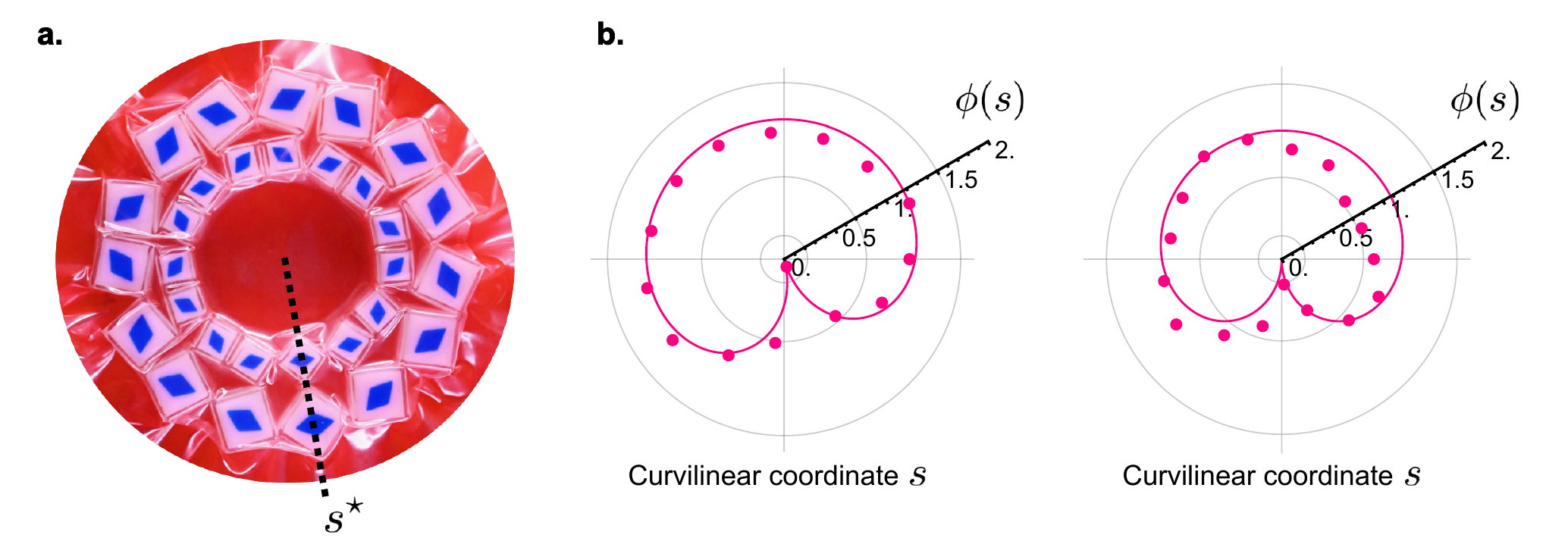}
\end{center}
\caption{{\bf Homogeneous compression of odd metachains}
{\bf a.} Picture of a metaring pressurized in a vaccuum bag ($N=15$) described in the Methods.
{\bf b.} Plots of the measured staggered rotation angle for two different rings of lengths $N=15$ and $N=17$ (symbols) and best fit from our continuum theory, Eq.~\eqref{eq.profileLongRibbon}. We measure a maximum deformation angle  $\varphi_{\rm max}=1.19\,\text{rad}$ and find $\xi_{15}=3.76\pm 0.60\, a_{15}$ and $\xi_{17}=2.21\pm0.13\, a_{17}$ where the lattice spacings are $a_{15}\approx12.6$mm and $a_{17}\approx 11.1$mm.
}
\label{fig.fitMetachain}
\end{figure}
We find the response  of odd metarings under the action of a homogeneous axial load by minimizing Eq.~\eqref{eq:phi4} with the constraint $\varphi(s^\star)=0$. This tedious minimization is discussed in~\cite{BartoloCarpentierPRX}. The solution corresponds to two symmetric half $\phi^4$ kinks continuously connected on the unit circle, see Fig.~\ref{fig.fitMetachain}. The analytical expression of the deformation profile takes a simple form only when expressed in an implicit form:
\begin{equation}
  \varphi(s)/\varphi_{\text{max}}=\pm\left|\sin\left( J\left[\frac{\sqrt{2\varphi_{\text{max}}}}{\xi}(s-s^\star),1\right]\right)\right|,
    \label{eq.profileLongRibbon}
\end{equation}
where $J[u,k]$ is the inverse of the elliptic integral function of the first kind, also known as the Jacobi Amplitude, and $\varphi_{\text{max}}$ is the maximum deformation angle.
This expression depends  on the metamaterial parameters only through the characteristic length $\xi$. We stress that this solution is singular at the location $s^\star$ and spontaneously breaks the translational invariance of Eq.~\eqref{eq:phi4}. As a result, the location of $s^\star$ is not prescribed and its translation along the base circle corresponds to  zero-energy  deformations. 
Fig.~\ref{fig.fitMetachain} reveals that our minimal theory expressed in term of a single scalar order parameter correctly accounts for our experimental findings. Eq.~\eqref{eq.profileLongRibbon} can therefore be effectively used to measure the elastic constant $K$ of pressurized metarings.

\section{Non-orientable mechanics of toroidal metamaterials}
We now consider a metatorus made of a square lattice of 2D lozenges with anti-ferromagnetic couplings  along both directions, Fig.~\ref{Fig:torus}a.
We note $N_{\rm p}$ and $N_{\rm t}$ the number of lozenges in the poloidal and toroidal directions. 
In this section, we first show  that all metatori including an odd number of units in at least one direction feature a non-orientable mechanics. 
We then demonstrate that non-orientability results in the topological protection of a zero-deformation loop under isotropic load.  
We finally show that the morphology of the zero-deformations loops are curves of minimal length that wind around the torus.
\subsection{Non-orientable deformation bundles and zero-deformation loops}
As in the Main Text, we consider the limit of very large tori and neglect all geometrical corrections arising from finite curvature.
In the long wave-length limit the elastic energy of a finite portion $U$ of the metatorus then generalizes Eq.~\eqref{eq:phi4} as:
\begin{equation}
    \mathcal E=\int_{U}\!\! \frac{K}{2}\left [\bm \nabla \varphi(x,y)\right]^2 +V(\varphi)\,{\rm d}x{\rm d}y,
    \label{eq:phi42D}
\end{equation}
where $x$ and $y$ are  the toroidal and poloidal coordinates, and $V$ is the  bistable quartic potential defined in Eq.~\eqref{eq:phi4}. 
Following the same analysis as for one-dimensional  rings, we  know that if number of lozenges is odd along  one, or two, directions then homogeneous anti-ferromagnetic order is frustrated.
We show below that this geometrical frustration  translates again in the non-orientability of the staggered-deformation bundle. 

\begin{figure}
\begin{center}
	\includegraphics[width=0.9\columnwidth]{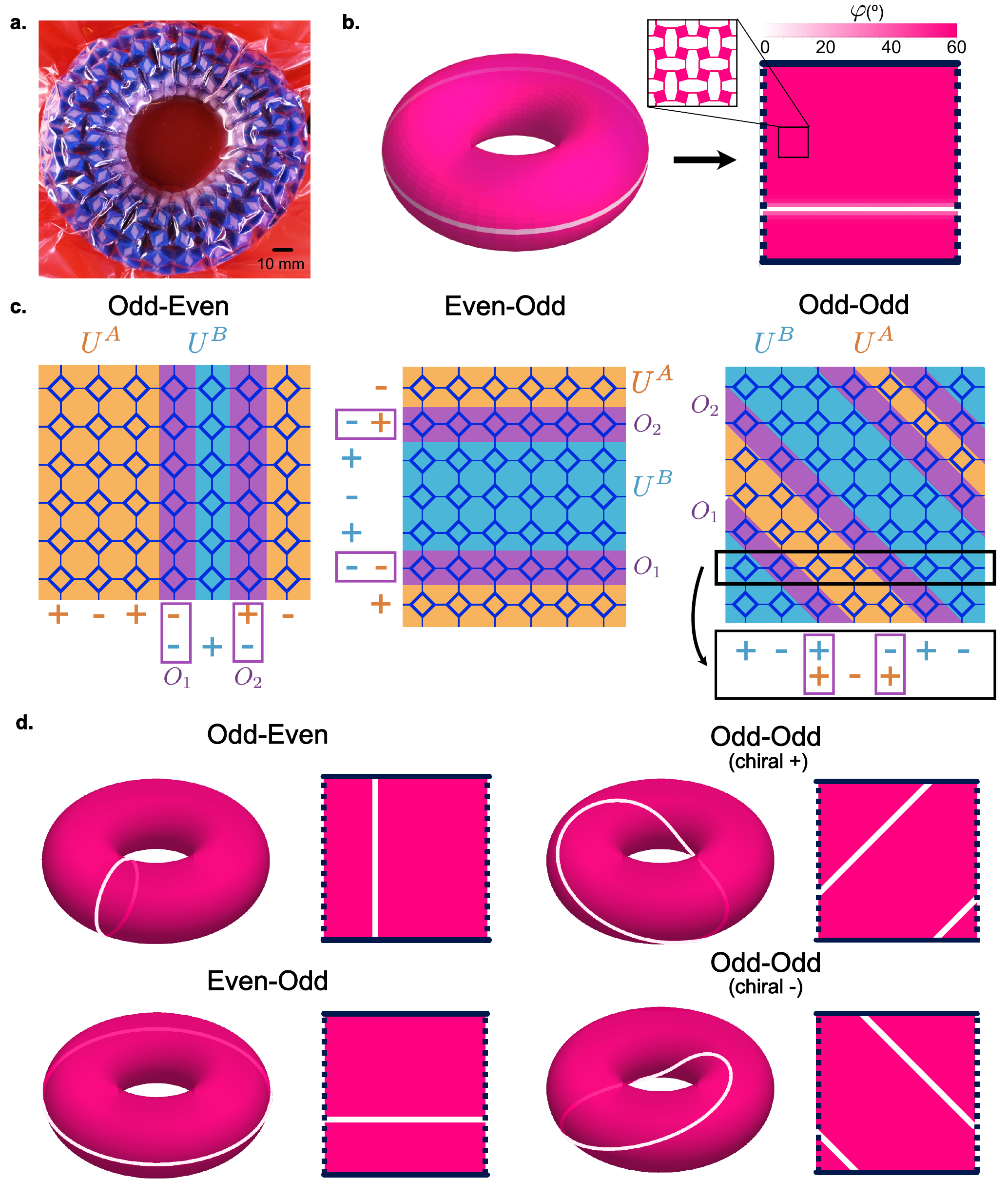}
\end{center}
\caption{{\bf Non-orientable mechanics of frustrated metatori.}
In all the figures $N_{\rm p}$ and,or, $N_{\rm t}$ are odd.
{\bf a.} Photograph of a toroidal metamaterial isotropically compressed in a vacuum bag ($N_{\rm p}=11$, $N_{\rm t}=27$).
{\bf b.} Planar representation of the torus. The heat map indicates the staggered deformation field on its planar representation, a square where the two pairs of opposite edges are identified. The deformations correspond to the minimization of the elastic energy Eq.~\eqref{eq:phi42D} with the constraint Eq.~\eqref{Eq.phiRstar} when $N_{\rm p}$ is odd and $N_{\rm t}$ is even. 
{\bf c.} Two overlapping open sets $U^A$ and $U^B$ cover $T^2$. The definition of  smooth staggered deformations on both $U^A$ and $U^B$ constrains the topology of the two sets. $U^{A/B}$ cannot include any closed loop winding along one direction where the number of lozenges is odd. We can always choose the orientation and the staggered-rotation fields $\varphi^{A/B}$ to coincide on the overlap region $O_1$. However, the parity of $N_{\rm p}$ and/or $N_{\rm t}$ forbids the equality $\varphi^{A}(x,y)=\varphi^{B}(x,y)$ when $(x,y)\in O_2$. No smooth anti-ferromagnetic order parameter having a constant sign can be defined on $T^2$. The staggered deformations define a non-orientable line bundle. 
{\bf d.} Topology and geometry of the zero-deformation line $\mathcal L$. When both $N_{\rm p}$ and $N_{\rm t}$ are odd  $\mathcal L$ winds along both directions and must spontaneously break two mirror symmetries. The shape of the $\mathcal L$ loops in the three plots correspond to the  minimum of the elastic energy. The loops are curved of minimal length which satisty the winding conditions set by the parity of $N_{\rm p}$ and $N_{\rm t}$.
}
\label{Fig:torus}
\end{figure}

For the sake of clarity, in Fig.~\ref{Fig:torus} we  represent the torus $T^2$ as a unit square with opposite edges identified, Fig.~\ref{Fig:torus}b.
We consider  two  open sets $U^A$ and $U^B$ that overlap and cover $T^2$. $U^A$ and $U^B$ are  associated to two  orientations $\epsilon^A$ and $\epsilon^B$, which  define two staggered rotation fields $\varphi^A(\mathbf r)$ and $\varphi^B(\mathbf r)$. 
 Any loop on the torus effectively realizes a 1D metaring. Hence when $N_{\rm t}$  is odd,  $U^{A}$ and $U^{B}$ should not include any loop winding in the toroidal direction. 
 Repeating the same reasoning we identify the three types of possible coverings sketched in Fig.~\ref{Fig:torus}b depending on the parity of $N_{\rm p}$ and $N_{\rm t}$.
Let us now assume that there exists a smooth staggered-rotation field in the limit $N_{\rm p},\,N_{\rm t}\gg1$ (keeping the parity of the number of lozenges unchanged). 
In the two overlaps  $O_1$ and $O_2$, the local rotation angle $\phi(x,y)$ is unambiguously defined with respect  to the normal of the torus. It is related to $\varphi^{A/B}$ as  $\phi(x,y)=\epsilon^A\varphi^A(x,y)=\epsilon^B\varphi^B(x,y)$ in the two overlap regions. 
We can always choose $\epsilon^A=\epsilon^B$ in  $O_1$  via a mere redefinition of the sign of $\varphi^B$.
A direct count of the number of sites separating $O_1$ and $O_2$ along the odd directions then implies that $\epsilon^A\varphi^A(x,y)=-\epsilon^B\varphi^B(x,y)$ in $O_2$.  
In other words,  there is an obstruction to define smooth  staggered deformations of constant sign over the whole torus.
The $\varphi(x,y)$ fields define a  non-orientable real line bundle $E\xrightarrow{\pi} T^2$~\cite{Gramain,Hatcher2002}. 

Generalizing the analysis done for one-dimentional rings in Sec~\ref{Sec:MetaRings}, we deduce that the largest open sets over which we can trivialize  the deformations is $T^2\backslash \mathcal L$, where $\mathcal L$ is a closed loop winding once around the torus, see Fig.~\ref{Fig:torus}d. 
As $E\xrightarrow{\pi} T^2$ is  non-orientable,  $\varphi$ must vanish along $\mathcal L$. 
We  conclude that the mechanics of frustrated toroidal metamaterials is defined by the combination of their elastic energy Eq.~\eqref{eq:phi42D}, and of the topological constraint:
\begin{equation}
\varphi\left(\mathbf R^\star(\sigma)\right)=0,
\label{Eq.phiRstar}
\end{equation}
where $\mathbf R^\star(\sigma)$ is a parametrization of $\mathcal L$ with respect to its curvilinear coordinate $\sigma$. 
The  deformation modes of the  $\mathcal L$ loop add a number of degrees of freedom to the staggered rotation variables. 
They must be dealt with when computing the equilibrium deformation patterns, which minimize the elastic energy Eqs.~\eqref{eq:phi42D} and~\eqref{Eq.phiRstar}. 
Remarkably,  without explicitly solving this constrained minimization problem, we already know that in the case where both $N_{\rm p}$ and $N_{\rm t}$ are odd, $\mathcal L$ must wind around both principal directions. 
As a result, mechanical equilibrium spontaneously breaks  two mirror symmetries: the emergent buckling patterns of odd-odd tori compressed by isotropic loads are chiral, Fig.~\ref{Fig:torus}c.

An alternative demonstration of the topological protection of the $\mathcal L$ loops can be done following the reasoning of Sec.~\ref{Sec.pedestrian}. 
Consider for instance the case where $N_{\rm t}$ is odd and $N_{\rm t}$ even.  
Starting from an anti-ferromagnetic lattice  model defined on two overlapping sets of sites, we can  take the continuous limit and find 
\begin{equation}
    {\mathcal E}_{\rm odd--even}=
    \int_{T^2} \left[\frac{K}{2} (\bm\nabla \varphi)^2
    +V(\varphi)\right]\,{\rm d}x{\rm d}y
    +\lim_{N_{\rm t},N_{\rm p}\to +\infty}4K~N_{\rm t}N_{\rm p}\int_{S_{1}}  \varphi^2(x^{\star}(y),y)\,{\rm d}y.
\end{equation}
A loop $\mathcal L$ lassoing the torus along the poloidal direction ($y$) is topologically protected against deformations irrespective of the magnitude of the load. 
$\mathcal L=\{\mathbf R^\star(y)\}$ is here parametrized in the Monge gauge: $\mathbf R^\star=(x^{\star}(y),y)$
%%%%%%%%%%%%%%%%%%%

We close this section by noting that the same analysis can be formally extended to higher dimensional tori $S_1^n$ where anti-ferromagnetic frustration results in the topological protection of vanishing deformations  on $n-1$ dimensional manifolds.

\section{Geometry of the zero-deformation loops}
\label{Sec:loop}
As described in the Main Text, when frustrated, toroidal  metamaterials are lassoed by zero-deformation loops $\mathcal L$. 
Their winding is determined by the underlying non-orientability  of the deformation bundle. The bundle topology, however, does not prescribe the loop geometry, which depends on the specifics of the load distribution and of the elastic energy. 
In this section, we show that under the action of isotropic loads $F$, the set of zero-deformation points form loops of minimal length and compute their associated line tension.

 Deep in the frustrated state, when $F\gg F_{\rm c}$, the rotation of the squares  are localized over a characteristic length $\xi=\sqrt{24Ka/F}$
 around $\mathcal L$.  
 This screening of the elastic deformations allows us to simplify the elastic energy. We first write  $(\bm\nabla \varphi)^2=(\partial_\parallel \varphi)^2+(\partial_\perp \varphi)^2$, 
 where $\partial_\parallel$ and $\partial_\perp$ are the derivatives in the directions longitudinal and normal to 
 $\mathcal L$.
 %$R^{\star}(\sigma)$. 
The large-scale variations of $\varphi$ then reduce to $(\bm \nabla \varphi)^2\approx \varphi_0^2/\xi^2$, %. 
where we recall that $\varphi_0$ is the amplitude of  the energy barrier separating the two minima of the quartic potential, Fig.~\ref{fig.SI1}. 
 The total energy $\mathcal E$ then takes the compact form:
 \beq
 \mathcal E \sim 2 \xi
 \int {\rm d}\sigma \left(\frac{K}{2}\frac{\varphi_0^2}{\xi^2}+\frac{K}{\xi^2}\varphi_0^4\right)\equiv \gamma \int {\rm d}\sigma,
 \label{Eq.gamma}
 \eeq
 with $\sigma$ the curvilinear coordinate along $\mathcal L$, and $\gamma=K\varphi_0^2(1+2\varphi_0^2)/(2\xi)$ the effective line tension.   
 In Eq.~\eqref{Eq.gamma}, we have effectively integrated out all the deformations  degrees of freedom $\varphi$. We are thus left with the residual degrees of freedom associated to the shape of the $\mathcal L$ loop. 
 We find that under a homogeneous load they are lines of minimal length topologically constrained to wind around the torus. Their three possible shapes are shown in Fig.~\ref{Fig:torus}d. 
 In the odd-even and even-odd cases $\mathcal L$ are circles (geodesics), whereas in the odd-odd case, $\mathcal L$ is chiral and forms a helix of minimal length, which lassos the torus both in the toroidal and poloidal directions. These predictions are in excellent agreement with our experimental findings reported in Fig.~2 (Main Text).
 
\section{Response to heterogeneous loads}
In this last section we detail the mechanical response of frustrated metamaterials under the action of heterogeneous loads. 
We first focus on one-dimensional rings including an odd number of units and show how  memory naturally emerges from non-orientable mechanics. 
We then compute the effective elasticity of the zero deformation loops arising from pointwise loads applied to metatori.
\subsection{Mechanical memory of one-dimensional metaring}
In the Main Text, we demonstrate the realisation of a  Set-Reset Latch memory by applying localised axial loads to odd metarings. 
Here we show that the possibility to read, write and erase mechanical bits stems from a multistable effective potential ruling the position $s^\star$ of the zero-energy node.

\begin{figure}
\begin{center}
	\includegraphics[width=1\columnwidth]{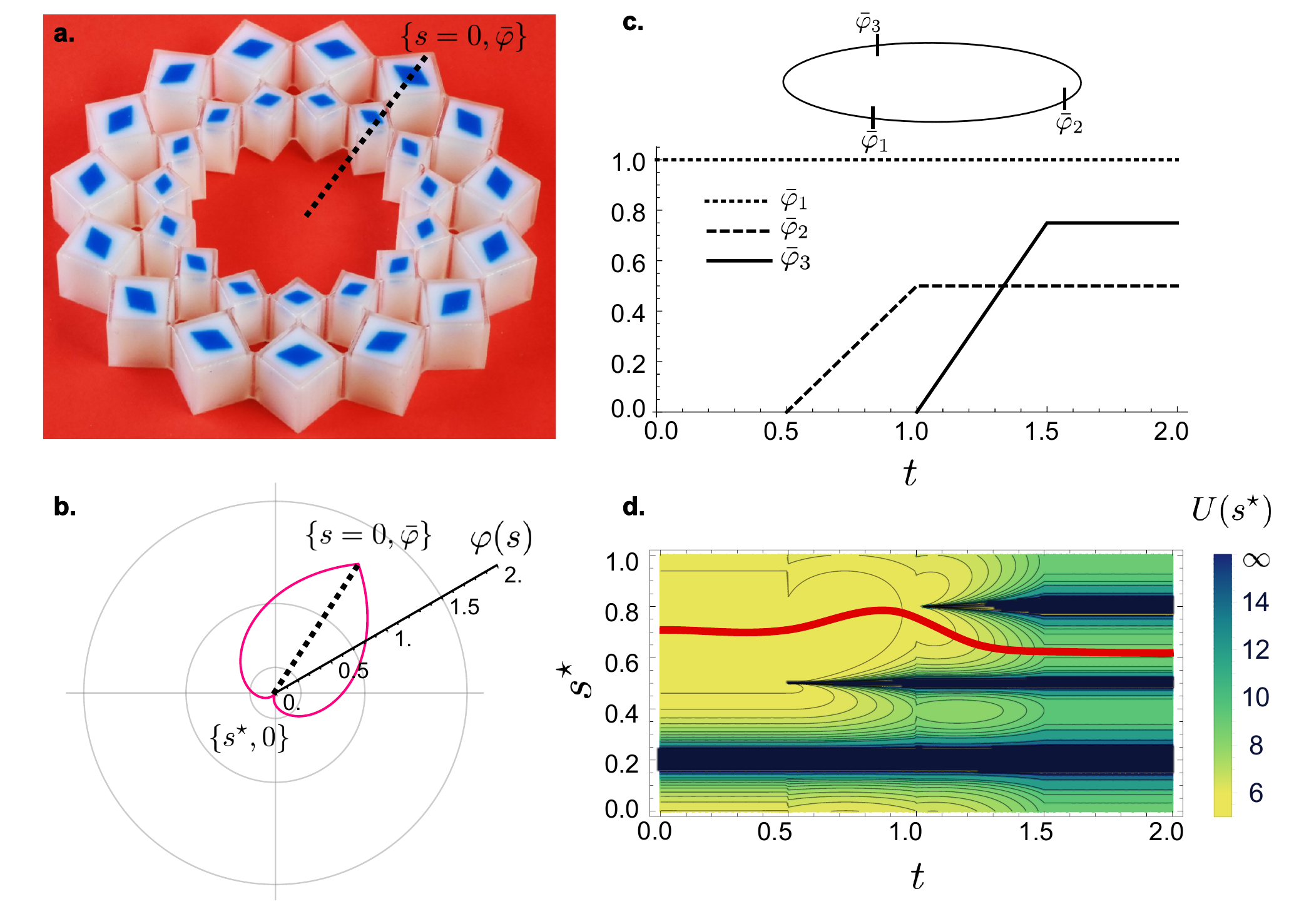}
\end{center}
\caption{{\bf Mechanical memory of one-dimensional metarings.}
{\bf a.} Response of a metaring (described in the Methods) to locally imposed deformations. Same experiments as in the Main Text.
{\bf b.} Polar plot of the  deformations induced by a point load located at $s_A$. The deformation vanishes at $s^\star$, the point opposite to $s_1$.
{\bf c.} Three local rotations are  imposed sequentially $(\bar\varphi_1, \bar\varphi_2,\bar\varphi_3$) at positions $s_1=0.2$, $s_2=0.5$, $s_3=0.8$, respectively, in a metaring consisting of an odd number of units.
{\bf d.} The  potential energy $U(s^\star)$ (heat map) evolves in time and becomes tri-stable when the three deformations are imposed.  The red curve indicates how the position of the zero-deformation node evolves in time as the deformations are sequentially applied.}
\label{fig.memory}
\end{figure}

We start from the continuum description of the odd metaring, and write the total energy
\begin{equation}
    \mathcal E=\int {\rm d}s\, \frac{K}{2}(\partial_s\varphi)^2+\frac{6C_b}{a^2}\varphi^2.
\end{equation}
We then readily find that the deformations of minimal energy satisfies
\begin{equation}
    \lambda^2\partial^2_{s}\varphi-\varphi=0,
    \label{eq.linearsystem}
\end{equation}
with $\lambda^2=a^2 K /(12C_b)$ and $\varphi(s^\star)=0$.
To investigate the response to an imposed deformation  at $s=0$. We add the extra constraint $\varphi(0)=\varphi(1)=\bar\varphi$ on the periodic field.
The deformation profile then takes the simple form:
\begin{equation}
    \varphi(s;s^\star)= \begin{cases} 
      A e^{s/\lambda}+Be^{-s/\lambda} & s\in[0,s^\star] \\
      C e^{s/\lambda}+De^{-s/\lambda}& s\in[s^\star,1] 
   \end{cases},
\end{equation}
with
\begin{align}
&\begin{pmatrix}A\\B\end{pmatrix}=\frac{\bar\varphi}{2\sinh(s^\star/\lambda)}\begin{pmatrix}-e^{-s^\star/\lambda}\\e^{s^\star/\lambda}\end{pmatrix},\nonumber\\
&\begin{pmatrix}C\\D\end{pmatrix}=\frac{\bar\varphi}{2\sinh((s^\star-1)/\lambda)}\begin{pmatrix}-e^{-s^\star/\lambda}\\e^{s^\star/\lambda}\end{pmatrix}.
\end{align}
We stress that the position $s^\star$ is yet to be determined. Inserting the deformation field $\varphi(s;s^\star)$ back in the elastic energy $\mathcal E$ leads to the effective potential:
\begin{equation}
    {\mathcal U}(s^\star)=\mathcal E[\varphi(s;s^\star)]=\frac{\bar\varphi^2}{2\lambda}\frac{\sinh (1/\lambda)}{\sinh((1-s^\star)/\lambda)\sinh(s^\star/\lambda)}.
\end{equation}
The location of the zero-energy node corresponds to the minimum of $\mathcal U$. 
When the ring is deformed by a single point load at $s=0$, $s^\star$ is diametrically opposed to the source of deformation, $s^\star=1/2$, see Figs.~\ref{fig.memory}a,~\ref{fig.memory}b.
The above result is readily generalised to a set of local deformations $\{\bar\varphi_n\}$ imposed at  $\{s_n\}$. To ease the notation we also include the topological restriction $\varphi(s^\star)=0$ as one of the imposed deformations. For each interval $[s_n,s_{n+1}]$, we have a solution $\varphi_n(s)=A_ne^{s/\lambda}+B_ne^{-s/\lambda}$, with
\begin{equation}
    \begin{pmatrix}A_n\\B_n\end{pmatrix}=M_n^{-1}\begin{pmatrix}\bar\varphi_n\\ \bar\varphi_{n+1}\end{pmatrix},\;\;\;\;\;\;
M_n=\begin{pmatrix}e^{s_n/\lambda} & e^{-s_n/\lambda}\\ e^{s_{n+1}/\lambda} & e^{-s_{n+1}/\lambda}\end{pmatrix}.
\end{equation}
As before, the determination of $s^\star$ follows directly from the minimisation of the effective potential ${\mathcal U}(s^\star)=\mathcal E[\{\varphi_n\}]$. If, for instance, $s_n=s^\star$ is located between $s_{n-1}$ and $s_{n+1}$, its exact position between both local deformations is given implicitly by
\begin{equation}
    \bar\varphi_{n+1}^2\sinh^2((s^{\star}-s_{n-1})/\lambda)=\bar\varphi_{n-1}^2\sinh^2((s^{\star}-s_{n+1})/\lambda).
\end{equation}
However, when more that one point load deforms the metamaterial $\mathcal U$ can feature as many minima as imposed deformations. This property lies at the core of the memory effect in odd metarings. As illustrated in Fig.~\ref{fig.memory}, the sequential application of local deformations leads to the programmed selection of a single minimum corresponding to one deformation pattern. 
This property echoes the mechanical response of non-orientable surfaces to shear stresses predicted in~\cite{BartoloCarpentierPRX}. Unlike in non-orientable surfaces, mechanical memory can be easily read given the planar geometry of the rings, and simply extended to higher dimensional meta-structures.

\subsection{Shape memory of the $\mathcal L$ loops.}

We now determine the shape of the $\mathcal L$ loop for frustrated metatori deformed by a collection of point stresses.
The intrinsic non-orientability of the staggered-rotation bundle $E\xrightarrow{\pi}T^2$ imposes the existence of zero deformation loops $\mathcal L$. 
When the applied load is homogeneous, elasticity minimises the length of $\mathcal L$, see Sec.~\ref{Sec:loop}. 
Under the action of localised sources of stress and deformations, the morphology of $\mathcal L$ is more difficult to anticipate. %
 In this last section,  we derive the effective elasticity of $\mathcal L$ loops caused by  heterogeneous distributions of localised loads and shed light on their intrinsic multistability. 

We consider a generic  load distribution  defined by the scalar field $F(\bm r)=\sum_\alpha F_\alpha 
\delta(\bm r - \bm r_\alpha )$. 
At mechanical equilibrium $\varphi$ minimises the total energy functional which takes the generic from
\beq
\mathcal E=\int d^2r \left(\frac{K}{2}(\nabla \varphi)^2+\frac{m^2}{2}\varphi^2\right)-\int d {\bm r} F(\bm r)\varphi.
\label{eq.energy}
\eeq
Non-orientability adds the topological constraint $\varphi(\bm R^{\star}(\sigma))=0$.
We stress that the minimisation of Eq.~\eqref{eq.energy} involves two coupled fields:(i) the staggered deformations,  $\varphi(\bm R^{\star})$, and the loop's shape $\bm R^{\star}(\sigma)$. 

To define the effective elasticity of $\mathcal L$,  we solve a seemingly more complex problem. We consider the case where the structure could also be subject to thermal fluctuations~\cite{Bartolo2003}, and introduce the partition function:
\beq
\mathcal Z=\int\mathcal D\varphi \int\mathcal D \bm R^{\star} ~ 
\left( \prod_{\sigma}\delta(\varphi({\bm R^{\star}}(\sigma)) \right) ~ e^{-\beta \mathcal E[\varphi,\bm R^{\star}]}.
\eeq
This formulation allows us to  compute the effective energy of the $\mathcal L$ loop by integrating out the the Gaussian fluctuations of $\varphi (\bm R^{\star})$. To proceed, we first express the topological 
constraint using a Lagrange multiplier 
$\lambda(\sigma)$. Noting,  $G[\bm r,\bm r']$  the Green function defined by 
$\beta(m^2-\kappa\nabla^2) G [\bm r,\bm r']=\delta(\bm r-\bm r')$, we have
\begin{align}
\mathcal Z&=\int \mathcal D \bm R^{\star} \int \mathcal D\lambda\int \mathcal D \varphi \ 
\exp\left[
    -\frac{1}{2}\int d\bm r d\bm r' ~\varphi(\bm r)G^{-1}(\bm r,\bm r')\varphi(\bm r')
    +\int d\bm r ~ F(\bm r)\varphi(\bm r)
    + {\rm i} \int d\sigma ~ \lambda(\sigma) \varphi(\bm R^{\star}(\sigma)) \right]\\
&=C_0 \int \mathcal D \bm R^{\star} \int \mathcal D\lambda 
    \exp\left[
    -\frac12 \int d\sigma ~d\sigma' ~ \lambda(\sigma) G[ \bm R^{\star}(\sigma)), \bm R^{\star}(\sigma')] \lambda(\sigma') 
    -{\rm i}\beta  \int d\sigma \int d^2\bm r ~ \lambda(\sigma) G[\bm R^{\star}(\sigma),{\bm r}]F(\bm r)\right]
\\
&= C_1 \int \mathcal D \bm R^{\star}  
    \exp\left( -{\mathcal U}[\bm R^{\star}] \right),
\label{eq:Zline0}
\end{align}
where $C_0,C_1$ are constants independent on $\bm R^{\star}(s)$ and the line effective 
energy in the presence of external forces reads 
\begin{equation}
    {\mathcal U}[\bm R^{\star}] = \frac{\beta^2}{2} \int d\sigma ~d\sigma' \int d^2\bm r d^2\bm r'\ 
    F(\bm r) G[{\bm r} ,\bm R^{\star}(\sigma)] \tilde{G}^{-1} [ \bm R^{\star}(\sigma), \bm R^{\star}(\sigma')]
    G[\bm R^{\star}(\sigma'),{\bm r}']F(\bm r'), 
\end{equation}
where $\tilde{G}^{-1}$ is defined by 
$\int d\sigma' G[ \bm R^{\star}(\sigma)), \bm R^{\star}(\sigma')]\tilde{G}^{-1}[ \bm R^{\star}(\sigma'), \bm R(\sigma'')]=\delta(\sigma-\sigma'')$. 
The case of a single force of amplitude $F_0$ applied at $\bm r_0$ corresponds to the 
simpler potential 
\begin{equation}
    {\mathcal U}_0[\bm R^{\star}] = \frac{\beta^2F_0^2}{2}  \int d\sigma ~d\sigma'  
G[{\bm r}_0 ,\bm R^{\star}(\sigma)] \tilde{G}^{-1} [ \bm R^{\star}(\sigma), \bm R^{\star}(\sigma')]
  G[\bm R^{\star}(\sigma'),\bm r_0].  
  \label{eq.exactPotential1source}
\end{equation}

Equation~\eqref{eq.exactPotential1source} although exact, is not handy since it requires the inverse of the non-local operator $\tilde G$. To gain more  insight on the long wave-length limit, we approximate the Green's function on the torus $G$ by its expression on the $\mathbb R^2$ plane: $G(\bm r,\bm r')=g(|\bm r-\bm r'|)=\frac{1}{2\pi \beta}K_0(m|\bm r-\bm r'|)$, with $K_n$ the modified Bessel function of the second kind.

\begin{figure}
\begin{center}
	\includegraphics[width=1\columnwidth]{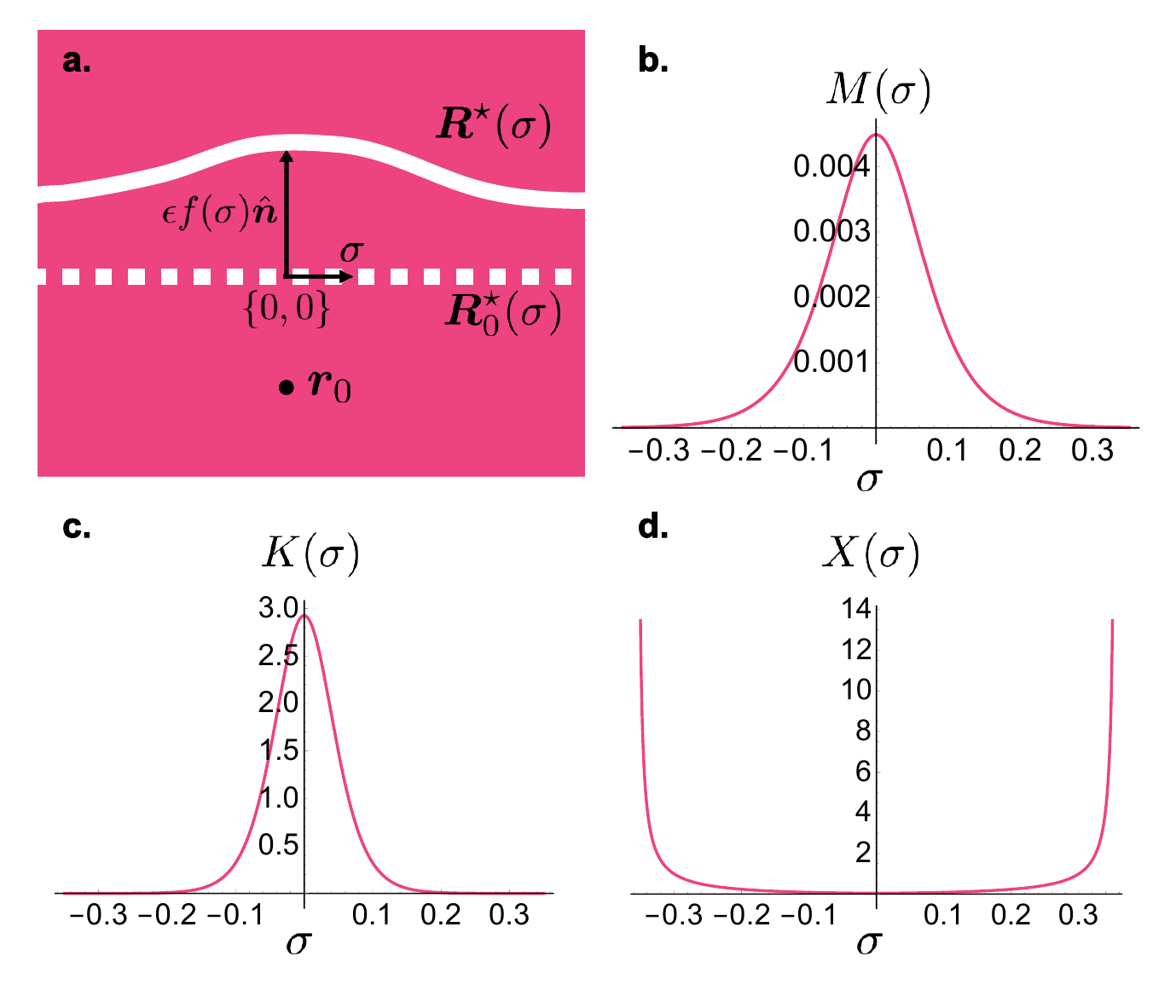}
\end{center}
\caption{{\bf Line repulsion from one point stress.}
{\bf a} Sketch of the unperturbed zero line $\bm R_0^{\star}(\sigma)=\hat{\bm t}\sigma+\bm p$ (dashed white line) and the deformed line $\bm R^{\star}(\sigma)=\bm R_0^{\star}(\sigma)+\epsilon f(\sigma)\hat{\bm n}$ (solid white line), where  $\hat{\bm t}=\{1,0\}$, and $\bm p=\{0,0\}$. The point stress is applied at $\bm r_0=\{0,-0.1\}$. {\bf b, c, d} Expansion terms from the one-source potential $\mathcal U_0$ for $m=10$: effective line tension $M(\sigma)$, repulsion term $K(\sigma)$, and shift term $X(\sigma)$, respectively.}
\label{fig.MKX}
\end{figure}

In response to a single point load, elastic energy is clearly minimized when placing $\mathcal L$ at a maximal distance from the load. 
This elementary reasoning is confirmed by 
our experiments and simulations. 
We hence  make the following ansatz for the loop shape $\bm R^{\star}_0(\sigma)=\hat{\bm t} \sigma+\bm p$, where $\hat{\bm t}$ is a unitary tangent vector, and $\bm p$ a point of reference.
Subject to additional local forces,  $\mathcal L$  undergoes transverse deformations. We parametrize its shape as $\bm R^{\star}(\sigma)=\bm R^{\star}_0(\sigma)+\epsilon \bm R^{\star}_1(\sigma)$, with $\bm R^{\star}_1(\sigma)=f(\sigma)\hat{\bm n}$, $\hat{\bm n}$ the normal vector, and $\epsilon$ the magnitude of the transverse fluctuation.

 To quadratic order in $\epsilon$ we have
\beq
G[\bm r_0 ,\bm R^{\star}(\sigma)]=G_0+\epsilon f(\sigma) G_1+\epsilon^2f^{2}(\sigma)G_2+\mathcal O (\epsilon^3)
\eeq
where
\begin{align}
    G_0[\bm r_0,\bm R^{\star}(\sigma)]&=g(|\hat{\bm t} \sigma+\bm p-\bm r_0|)\nonumber\\
    G_1[\bm r_0,\bm R^{\star}(\sigma)]&=\frac{\hat{\bm n}\cdot(\bm p-\bm r_0)}{|\hat{\bm t} \sigma+\bm p-\bm r_0|}g'(|\hat{\bm t} \sigma+\bm p-\bm r_0|)\nonumber\\
    G_2[\bm r_0,\bm R^{\star}(\sigma)]&=\left[\left(\frac{|\hat{\bm t} \sigma+\bm p-\bm r_0|^2-(\hat{\bm n}\cdot(\bm p-\bm r_0))^2}{4|\hat{\bm t} \sigma+\bm p-\bm r_0|^3}\right)g'(|\hat{\bm t} \sigma+\bm p-\bm r_0|)+\frac{(\hat{\bm n}\cdot(\bm p-\bm r_0))^2}{4|\hat{\bm t} \sigma+\bm p-\bm r_0|^2}g''(|\hat{\bm t} \sigma+\bm p-\bm r_0|)\right].
\end{align}
Analogously we have
\begin{equation}
    \tilde G[\sigma,\sigma']=\tilde G_0+\epsilon\tilde G_1+\epsilon^2\tilde G_2+\mathcal O(\epsilon^3),
\end{equation}
with
\begin{align}
    \tilde G_0[\sigma,\sigma']&=g(|\sigma-\sigma'|)\nonumber\\
    \tilde G_1[\sigma,\sigma']&=\hat{\bm t}\cdot\hat{\bm n}\frac{(f(\sigma)-f(\sigma'))(\sigma-\sigma')}{|\sigma-\sigma'|}g'(|\sigma-\sigma'|)=0\nonumber\\
    \tilde G_2[\sigma,\sigma']&=\frac{(f(\sigma)-f(\sigma'))^2}{2|\sigma-\sigma'|}g'(|\sigma-\sigma'|).
\end{align}

Notice that the operator $\tilde G$ is non-local. In order to have a local potential $\mathcal U$ we employ the last approximation: $m l\gg1$, where $l$ is the characteristic separation between two neighboring sites. Under this regime, $g(\sigma)\sim\delta(\sigma)$ rendering both $\tilde G$ and $\tilde G^{-1}$ local functions of $\sigma$.

All in all, after eliminating unnecessary constant shifts in energy, the potential of eq.~\eqref{eq.exactPotential1source} reads
\begin{equation}
{\mathcal U}_0[\bm R^{\star}]=\epsilon \int d\sigma f(\sigma)K(\sigma)X(\sigma)+\epsilon^2\int d\sigma\left[\frac{M(\sigma)}{2}\left(\frac{df}{d\sigma}\right)^2+\frac{K(\sigma)}{2}f^2(\sigma)\right], \label{eq:U0final}   
\end{equation}
where
\begin{align}
    M(\sigma)&=G_0^2,\nonumber\\
    K(\sigma)&=2\left(G_1^2+2G_0G_2\right),\nonumber\\
    X(\sigma)&=-\frac{G_1G_0}{G_1^2+2G_0G_2},\nonumber
\end{align}
and the arguments of the $G$ function are implicitly given by $[\bm r_0,\bm R^{\star}(s)]$. These three functions are plotted in Fig.~\ref{fig.MKX}

We can now qualitatively understand the response of the $\mathcal L$ loop to a point-wise stress applied to the metatori. 
At first order in $\epsilon$, Eq.~\eqref{eq:U0final} indicates that the repulsion from the source of stress is minimised when the loop is at a maximal distance from the source. $\mathcal L$ loops are repelled by point stresses.
At second order in $\epsilon$, Eq.~\eqref{eq:U0final} translates the effective elasticity of the $\mathcal L$ loop. The first term of order $\epsilon^2$ acts as an effective line tension. Unlike in the case of homogeneous load, the line tension is here heterogeneous. The second and last term of order $\epsilon^2$ modulates the repulsive force from the source.

Simply put $\mathcal L$ loops have a finite line tension and are repelled from point stresses. These two ingredients shed light on the pattern dynamics showed in Fig. 3 in the Main Text, where we see the zero deformation loop moving and bending away from the applied perturbations.

\section{Experimental Methods}

\subsection{Sample designs}

\begin{figure}[b!]
    \centering
    \includegraphics[width=0.8\textwidth]{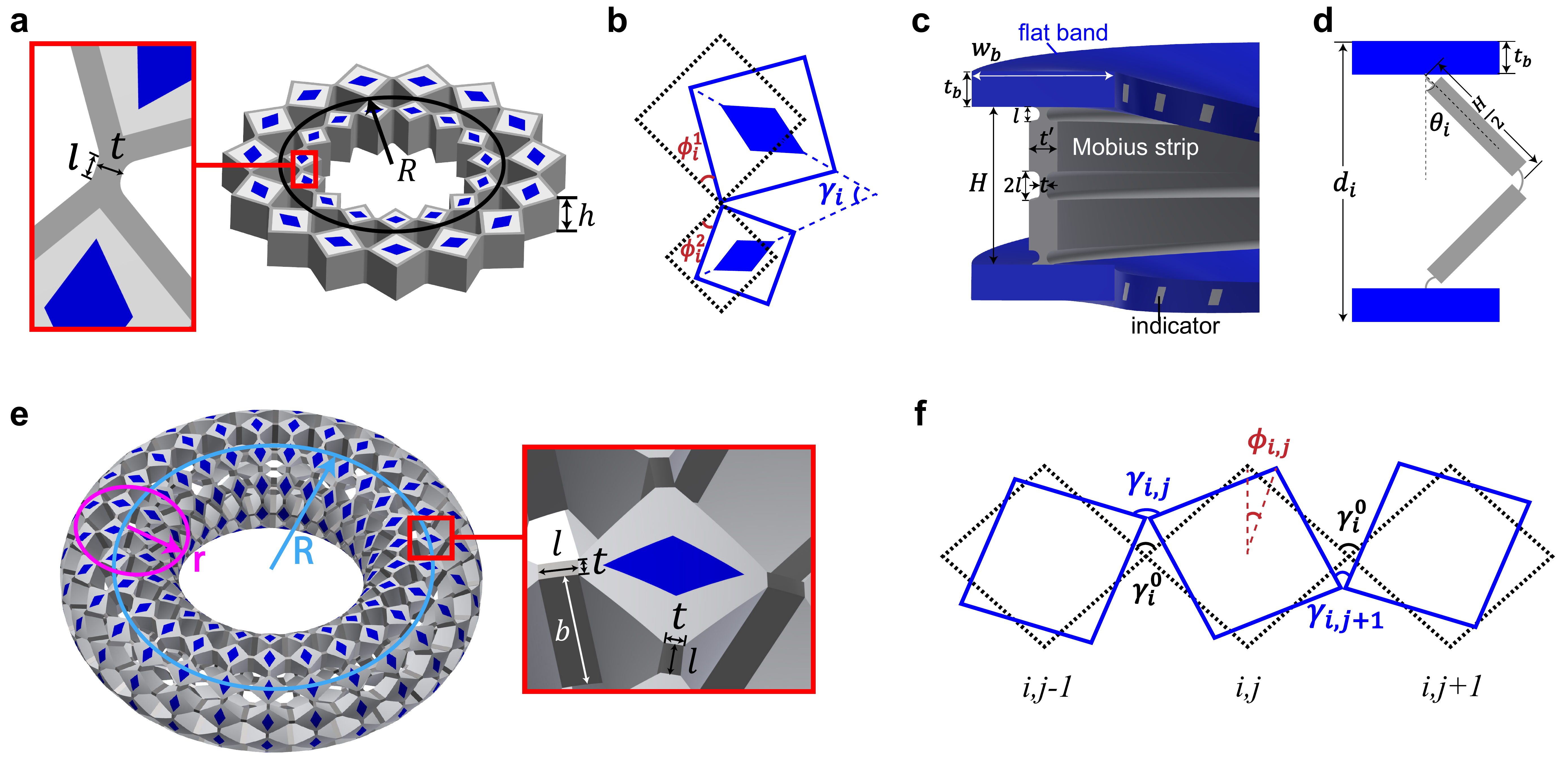}
    \caption{\textbf{Geometry and data acquisition.}{ \textbf{(a)} Geometry design of a metaring, where $R=30$ mm, $h=12$ mm, $t=1$ mm and $l\approx1$ mm. Squares are marked by diamond marks to track their position and rotation. \textbf{(b)} Definition of the rotation angle of a metaring's unit cell  $\phi_i=(\phi^{1}_i+\phi^{2}_i)=\gamma_i$. \textbf{(c)} Cross section of twisted bands, where $w_b=20$ mm, $t_b=5$ mm, $H=22$ mm, $t=1$ mm, $l=2$ mm and $t'=4$ mm. The radius of twisted bands is 50 mm. A coating of flexible material (Stratasys Agilus30) encloses the flat band in order to make the junction more durable. \textbf{(d)} Bending angle of a twisted band $\theta_i=\arccos{(({d_i-2t_b})/{H})}$.  \textbf{(e)} Geometry design of toroidal metamaterials, where $R=51$ mm, $r=20$ mm, $l=2$ mm, $t=1$ mm and $b=8$ mm. \textbf{(f)} Definition of the rotation angle of a metatorus' unit cell $\phi_{i,j}=\frac{1}{2}[( \gamma_{i,j}-\gamma^{0}_{i})-( \gamma_{i,j+1}-\gamma^{0}_{i})]$. A coating of flexible material (Stratasys Agilus30) encloses the squares in order to make the hinge more durable.}
    }
    \label{Ex_fig_geometry_design}
\end{figure}

\paragraph{Metarings}
We design metarings, formed of 15, 17, 16 and 18  pairs of squares connected by short hinges connecting their corners, see Fig.~\ref{fig:1}a and d \mtf~and Fig.~\ref{Ex_fig_geometry_design}a. The high contrast in stiffness between the squares and the hinges and the geometry of the soft hinges ensure that the shearing stiffness is larger than the bending stiffness. 
This implies that counter-rotations of adjacent squares is more energetically favourable than co-rotation~\cite{Coulais_NatPhys2017}. In other words antiferromagnetic order develops over long distances.

\paragraph{Flexible twisted bands}
We design four types of twisted bands that have 0, 1, 2 and 3 half twists and that form closed loops, see Fig.~\ref{fig:1}f and i \mtf. 
Since we are interested in their out-of-plane buckling under homogeneous compression, we design the cross section of the twisted band in the shape of the  ``I'' letter, see Fig.~\ref{Ex_fig_geometry_design}c. 
The top and bottom parts of the I are made of a stiff material and are used for applying compressive loads. The middle part of the twisted band is made of a soft material and its dimensions are suitably designed such that it features a single buckling mode.

\paragraph{Metatori}
The geometry of the metatori is that of  toroidal shells formed of a square lattices of lozenges connected via their corners, see Fig.~\ref{fig:2}a,d and g \mtf~and Fig.~\ref{Ex_fig_geometry_design}e. We design three types of metatori. They correspond to lattices including numbers of lozenges in the toroidals and poloidal directions with all possible parities. 
We again use a stiff material for the lozenges and a soft material for the hinges. This design promotes antiferromagnetic order under compression.

\subsection{Sample fabrication}
\begin{figure*}[b!]
    \centering
    \includegraphics[width=0.5\textwidth]{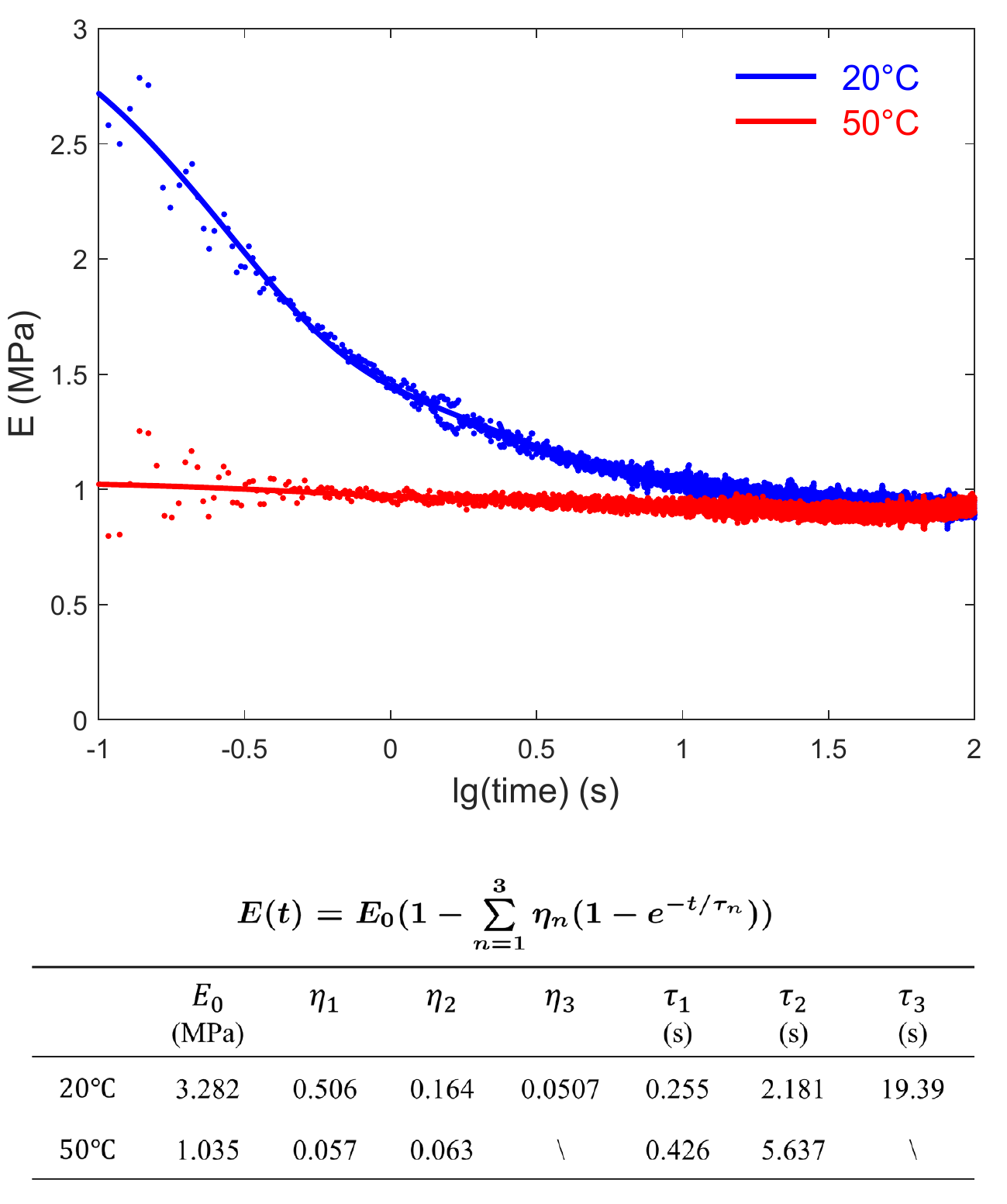}
    \caption{\textbf{High temperature weakens the viscoelasticity of Agilus.} The sample was stretched under a high strain rate of 4/s up to strain $\epsilon=0.2$ and held fixed for 100 s. The effective section of the standard sample has a length $L=80$ mm, depth $d=4$ mm and width $w=20$ mm. The Young's modulus is obtained by the neo-Hookean material model, $E(t)=3F/[(\lambda-\lambda^{-1})wd]$, where $\lambda$ is the applied stretch ratio and $F$ is the measured force. The solid lines are fits of the Maxwell–Wiechert viscoelastic material model to the data, given by $E(t)=E_0(1-\sum\limits_{n=1}^3\eta_n(1-e^{-t/\tau_n}))$, where $E_0$ is the peak Young’s modulus under instantaneous load, $\eta_n$ is the dimensionless relaxation strength and $\tau_n$ is the timescale. At room temperature ($20^\circ$C), the effective Young’s modulus drops by $70\%$ in $20 s$. While at $50^\circ$C, the effective Young’s modulus only drops by up to $11\%$.} 
    \label{Ex_fig_agilus_temperature}
\end{figure*}

We produce all the samples by additive manufacturing using a PolyJet 3D printer (Stratasys Object500 Connex3). The central parts of the twisted bands and of the hinges of metarings and metatori are made of a flexible photopolymer (Stratasys Agilus30, Young's modulus $E\approx1$~MPa, see Fig.~\ref{Ex_fig_agilus_temperature} for a calibration of the elastic properties). The other parts, the top and bottom part of the twisted bands and the lozenges forming the metarings and metatori, are made of a stiffer material (Stratasys Vero, Young's modulus $E\approx2500$~MPa).

\subsection{Experimental setup}
\paragraph{Homogeneous pressure}
In Fig.~\ref{fig:1} and Fig.~\ref{fig:2} \mtf, we test our structures under homogeneous compression. To this end, the samples are placed into a seal plastic bag connected to a vacuum pump (KNF Neuberger N022.AN.18) and a vacuum regulator (SMC IRV series). 
We use this device to apply a pressure difference of $\sim20$~kPa, such that the bag applies a constant pressure on the structures.

\begin{figure*}[t!]
    \centering
    \includegraphics[width=0.7\textwidth]{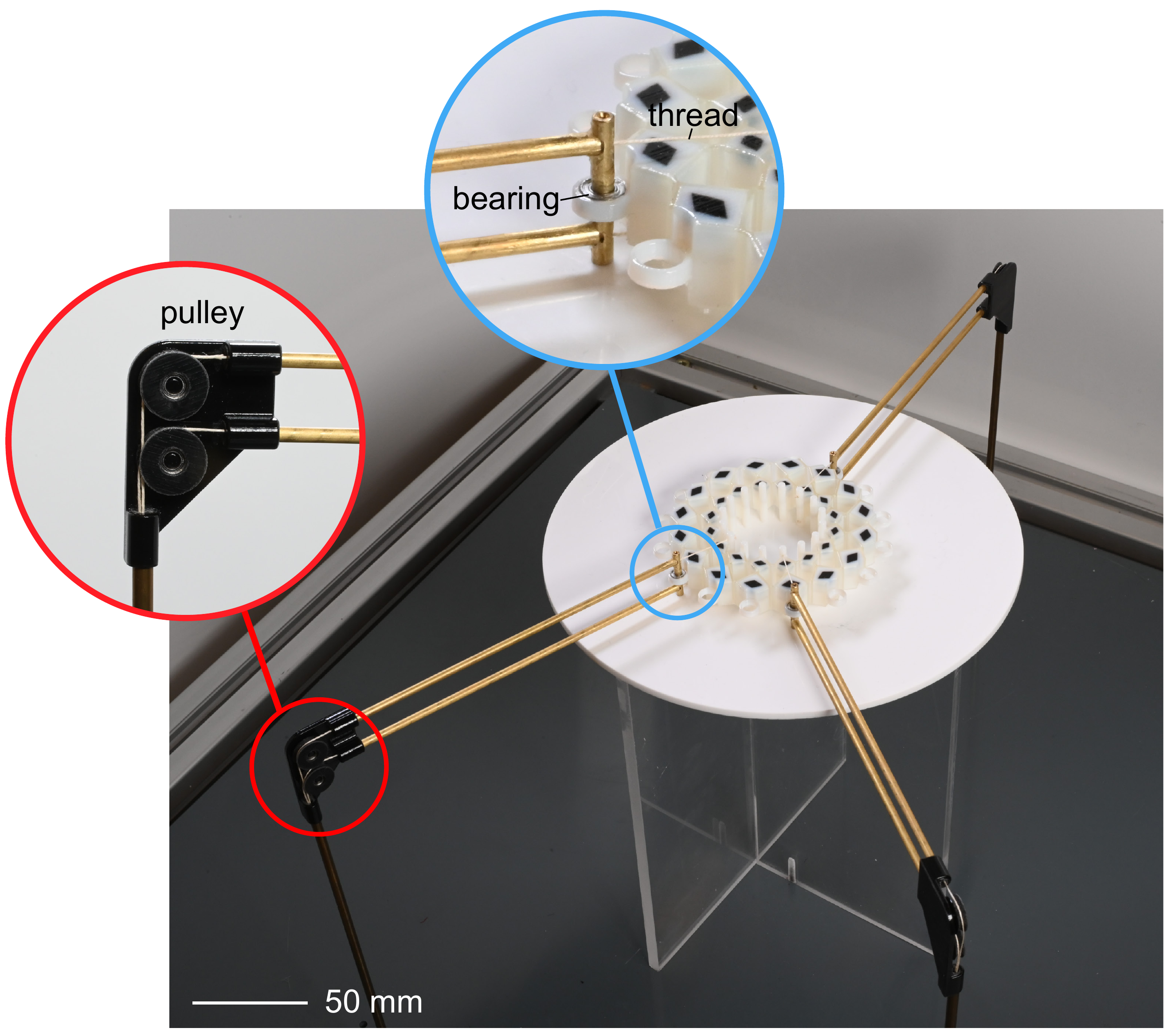}
    \caption{\textbf{Setup of applying local forces on a metaring.}}
    \label{Ex_fig_setup}
\end{figure*}

\paragraph{Local compressive loads}
To demonstrate  non-Abelian mechanics and the mechanical sequential logic gates displayed in Fig.~\ref{fig:3} \mtf~and in \EDF~\ref{Ex_fig_non_abelian_curve}, we designed a second setup shown in Fig.~\ref{Ex_fig_setup}. The setup allows us to apply local compressive forces using pulleys and weigths. 
To minimize the effect of unwanted external torques that would bias our measurements, we attached ball bearings to the metaring. 
In the non-Abelian mechanics experiments (Fig.~\ref{fig:3}b \mtf~ and \EDF~\ref{Ex_fig_non_abelian_curve}), 
we use an odd (even) metaring made of 15 (16) pairs of squares, and apply three forces P, S and R having the same magnitude (2.75 N). The point of applications of the loads are $i_P=1$, $i_S=7$ and $i_R=11$ ($i_P=1$, $i_S=8$ and $i_R=12$).

In the mechanical Set-Reset latch experiment  (Fig.~\ref{fig:3}d \mtf), we use the same points of applications of the three loads P, R, S as above, but we slightly adjust the experimental protocol to ensure that the hinges can snap back during the unloading steps: (i) we use smaller P, S and R loads of 1.65N, 1.05N and 1.65N; (ii) we work at a temperature of $50^\circ$C by using a heat gun and a thermometer (Fluke 62 Mini). At room temperature, the flexible photopolymer (Agilus30) features a strong viscoelastic stress relaxation, which typically prevents snap back~\cite{Dykstra_JAM2019}. As  shown in Fig.~\ref{Ex_fig_agilus_temperature}, a temperature of $50^\circ$C speeds up the viscoelastic stress relaxation by orders of magnitude and limit the fluctuations of the Young modulus .

\subsection{Data acquisition}

\paragraph{Metarings}
The experiments on the metarings are recorded with a high-resolution camera (Nikon D780 with Micro-Nikkor 105mm lens, resolution $6048$~px~$\times$~$4024$~px). By tracking the diamond markers on the squares using particle detection and tracking (ImageJ), we measure the position and angle of each square. The accuracy of the measured rotation angles is $0.3$~deg. From the rotation of each square, we extract the average rotation of the pairs of squares $i$ $\phi_i=(\phi^{1}_i+\phi^{2}_i)$, see Fig.~\ref{Ex_fig_geometry_design}b, from which we compute the staggered variable $\varphi_i$. The data is shown in Fig.~\ref{fig:1}d and e \mtf, Fig.~\ref{fig:3}b and d \mtf, \EDF~\ref{Ex_fig_non_abelian_curve} and in Supplementary Videos~2, 5 and 6.  

\paragraph{Flexible twisted bands}
The distances $d_i$ between the white square markers on the deformed flexible twisted bands are measured by a digital caliper (Mitutoyo CD-15APX). The buckling angle is then computed from the distances $d_i$ and from the dimensions of the specimen using the formula $\theta_i=\arccos{(({d_i-2t_b})/{H})}$, see Fig.~\ref{Ex_fig_geometry_design}d. Error propagation on all these measurements give a typical measurement error of 5 degrees for the buckling angle. The data is used in Fig.~\ref{fig:1}i and j \mtf.

\paragraph{Metatori}
The data from the experiments of toroidal metamaterials correspond to measurements performed with a digital protractor (Wixey WR41). The error is around 5 degrees. By measuring the angle $\gamma_{i,j}$ between adjacent lozenges, the rotation of each square $\phi_{i,j}=\frac{1}{2}[( \gamma_{i,j}-\gamma^{0}_{i})-( \gamma_{i,j+1}-\gamma^{0}_{i})]$ is obtained, see Fig.~\ref{Ex_fig_geometry_design}f. The data are plotted in Fig.~\ref{fig:2}b, e, and h \mtf~and shown in Supplementary Video 4.

\subsection{Numerical simulations}
The numerical results shown in Fig.~\ref{fig:3}e \mtf~and in the Supplementary Video 7 correspond to Brownian dynamics simulations over the angle variables $\phi_{i,j}$ indexed by the coordinates $i,j$ of the sites:
\begin{equation}
\eta\frac{d\phi_{i,j}}{dt}=-\sum_{i’,j’}\mathcal D_{i,j,i’,j’}\phi_{i’,j’}-f_{i,j}(t)\phi_{i,j}\left(\phi_{i,j}^2-\phi_0^2\right)+w(t),
\label{eq.BrowEq}
\end{equation}
where $\eta$ is the friction coefficient, $f_{i,j}(t)$ corresponds to the force applied at the site $i,j$ at a time $t$, $w(t)$ is a white noise signal of small amplitude (0.01), and the dynamical matrix $\mathcal D$ is directly obtained as the Hessian of the discrete elastic energy
\begin{align}
    E=&\sum_{i,j}\left[\frac{C_s}{2a}(\phi_{i+1,j}+\phi_{i,j})^2+\frac{C_s}{2a}(\phi_{i,j+1}+\phi_{i,j})^2\right]\nonumber\\
    &+\sum_{i,j}\left[\frac{C_b}{2a}(\phi_{i+1,j}-\phi_{i,j})^2+\frac{C_b}{2a}(\phi_{i,j+1}-\phi_{i,j})^2\right],
\end{align}
satisfying $\mathcal D_{i,j,i',j'}=\partial E/\partial\phi_{i,j}\partial\phi_{i',j'}$.
The white noise is used to ensure that a stable equilibrium is reached but none of the results depend on its magnitude.

Eq.~\eqref{eq.BrowEq} was solved with the software \textit{Mathematica} using a LSODA scheme, which automatically detects stiff and non-stiff regions, adapting the solver as needed~\cite{hindmarsh1983odepack}.

The simulations were performed on systems with periodic boundary conditions of sizes 48x49, 49x48, and 49x49 for the even-odd, odd-even, and odd-odd cases respectively. 

The simulations involving a homogeneous compression of the metatori (Fig. \ref{Fig:torus}) were performed by setting $\eta=2$, $a=1$, $\phi_0=1$, $C_s=3$, $C_b=0.01$, and $f_{i,j}(t)=1\;\;\forall i,j,t$ and a total simulation time $T=100$. The choice of bending and shearing stiffness ensures a strong anti-ferromagnetic order ($C_s/C_b\gg1$).

For the localised loading case shown in Fig.~\ref{fig:3}e \mtf, we used $\eta=1$, $a=1$, $\phi_0=1$, $C_s=1$, $C_b=0.001$, and the magnitude of the loading forces were kept constant in each simulation: $|f^{\text{t}}_{i,j}(t)|=70$. We used a total simulation time of $T=1050$.  As initial conditions we used configurations with predefined zero-deformation lines. We then let the system relax for a time $t^{\text{relaxation}_1}=30$ in both cases, reaching an equilibrium configuration (bottom row of each subfigure). Then the applied sources are braided in as follows: exchange of local loads for a duration of $\tau=400$, relaxation of  $t^{\text{relaxation}_2}=20$, new exchange of $\tau=400$ and a final relaxation of $t^{\text{relaxation}_3}=200$.

\subsection{Non-Abelian conditions}
In odd metarings, non-Abelian mechanical responses are achieved by controlling the path of the zero node upon different loading sequences. Crucially, non-Abelian response does not occur generically and the points of applications of the three forces must obey specific constraints. 

\begin{figure*}[t]
    \centering
    \includegraphics[width=\textwidth]{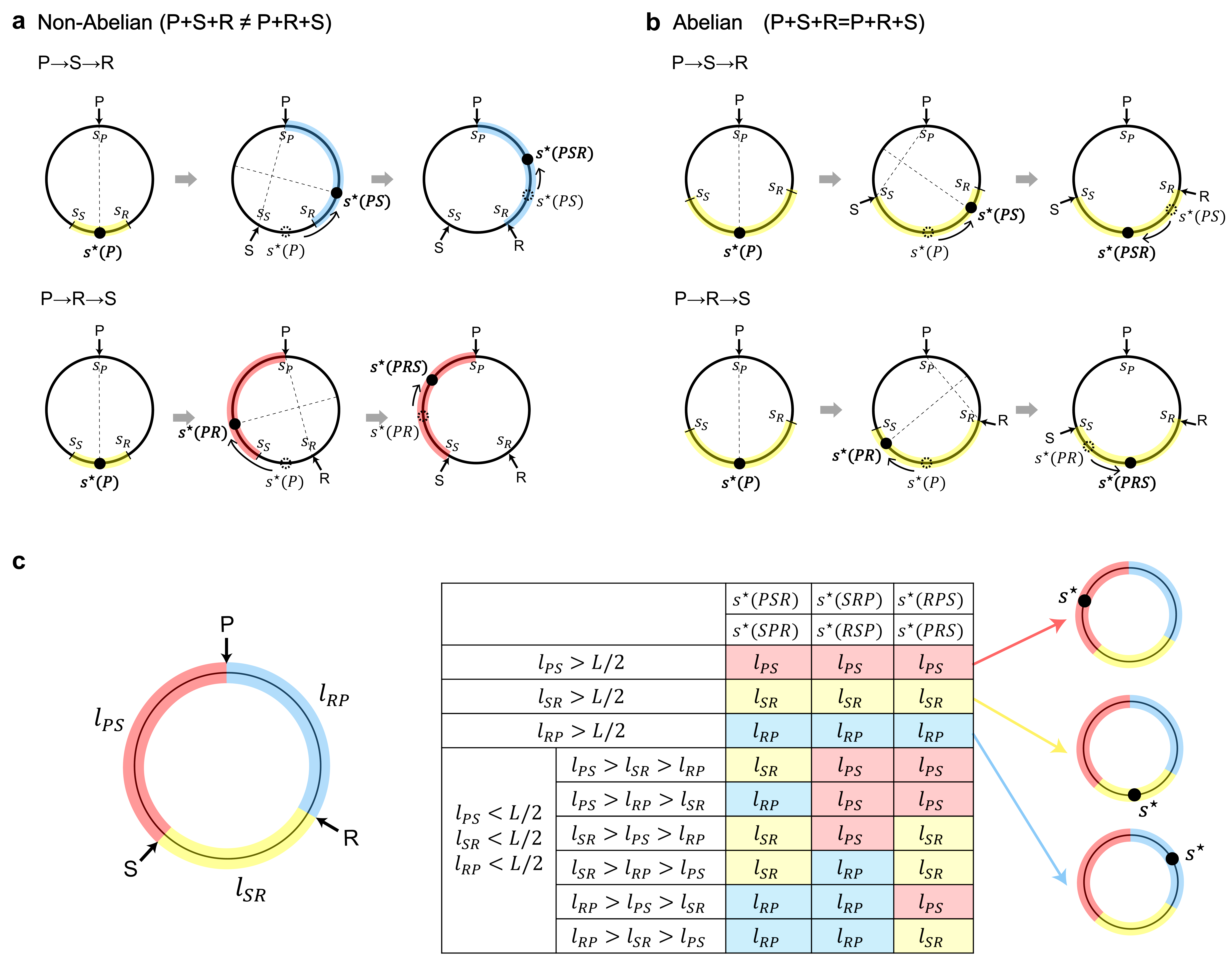}
    \caption{\textbf{Conditions of non-Abelian mechanics of an odd metaring under three local forces.} {\textbf{(a)} A case where the points of applications of the loads P, R, S leads to non-Abelian mechanics. Under two different loading sequences, the third force pushes the zero nodes $s^{\star}$ towards distinct sectors. \textbf{(b)} A case where the points of applications of the loads P, R, S do not allow for a non-Abelian response. Under two different loading sequences, the third force push the zero nodes $s^{\star}$ back towards the same sector. \textbf{(c)} Conditions for non-Abelian responses based on the relative sizes of the sectors $l_{PS}$, $l_{SR}$ and $l_{RP}$, and on the sequential order of the loads P, R, and S and location of zero nodes $s^\star$ resulting from  six loading sequences. The zero node can be located at three possible positions, which are respectively on the sector $l_{PS}$, $l_{SR}$ and $l_{RP}$. $L$ is the length of the whole loop. }
    }
    \label{Ex_fig_non_abelian_conditions}
\end{figure*}

The $N$ point loads compartmentalise the loop into $N$ sectors. Non-Abelian response is achieved if, at the end of two different loading sequences, the zero nodes end their journeys in two different sectors, see Fig.~\ref{Ex_fig_non_abelian_conditions}a. Conversely, the response is Abelian if, at the end of two different loading sequences, the zero nodes end their journeys in the same sector, see Fig.~\ref{Ex_fig_non_abelian_conditions}b.

As we show in the SI, the location of the zero node $s^\star$ can be calculated by minimising the energy density Eq.~(\ref{eq:elastic}) \mtf~under the constraint $\varphi(s^\star)=0$. This calculation also predicts that, in the case of equal loads, the zero node $s^\star$ moves as far as possible from the points of application of the loads. This basic repulsive interaction between the loads and the zero nodes allows us to derive the conditions for non-Abelian response by using a purely geometric argument. This argument applies generically for an arbitrary number of $N$ loads. We use this geometric approach to derive the conditions for non-Abelian response in the case of three loads, see Fig.~\ref{Ex_fig_non_abelian_conditions}c. These conditions translate into inequalities for the relative size of the $3$ sectors and prescribe which specific sequential order of the load applications moves the zero node $s^*$ in a specific  sector. Importantly, although  in principle three possible states can be achieved for given positions of the load and sizes of the sectors, only two possible states can be reached by using different sequences. We have used these considerations to determine the points of applications of the loads for the experiments in Fig.~\ref{fig:3}b and d \mtf~and in \EDF~\ref{Ex_fig_non_abelian_curve}.

  \end{document}